\documentclass[aps,prd,twocolumn,superscriptaddress,nofootinbib]{revtex4-2}
\usepackage{graphicx}
\usepackage{breqn}
\usepackage{mathtools}
\usepackage{amsmath}
\usepackage[dvipsnames]{xcolor}
\usepackage{xcolor}
\usepackage{aas_macros}

\usepackage{mathrsfs}

\usepackage{tablefootnote} 
\usepackage{tabularx}
\usepackage{subfig}
\usepackage{booktabs}
\usepackage{url}
\usepackage{soul}
\newcommand{\hatbf}[1]{\mathbf{\hat{#1}}}

\newcommand{\mg}{\mathrm{MG}}
\newcommand{\gr}{\mathrm{GR}}
\newcommand{\hatVG}{\widehat{V}_G}
\newcommand{\vgCl}{\widetilde{C}_{\ell}^{vg^{\dagger}}}
\newcommand{\ggCl}{{C}_{\ell}^{gg}}
\newcommand{\kgCl}{{C}_{\ell}^{\kappa g}}

\newcommand{\zeff}{z_{\mathrm{eff}}}
\newcommand{\chieff}{\chi_{\mathrm{eff}}}
\newcommand{\Keff}{K_{\mathrm{eff}}}
\newcommand{\kvec}{\textbf{k}}
\newcommand{\potdiff}{\nabla^2 (\psi - \phi)}
\newcommand{\gdag}{g^{\dagger}}
\newcommand{\vrad}{\hat{v}_r}
\newcommand{\mur}{\mathrm{cos}(\vartheta)}
\newcommand{\dndz}{\frac{\mathrm{d}N}{\mathrm{d}z}}

\graphicspath{{./}{figures/}}

\begin{document}
\title{Probing Gravity at Large Scales with kSZ-Reconstructed Velocities and CMB Lensing} 
\author{Raagini Patki}
\email[Corresponding author:\ ]{rp585@cornell.edu}
\affiliation{Department of Astronomy\char`,{} Cornell University\char`,{} Ithaca\char`,{} NY 14853\char`,{} USA.}
\author{Nicholas Battaglia}
\affiliation{Department of Astronomy\char`,{} Cornell University\char`,{} Ithaca\char`,{} NY 14853\char`,{} USA.}
\author{Rachel Bean}
\affiliation{Department of Astronomy\char`,{} Cornell University\char`,{} Ithaca\char`,{} NY 14853\char`,{} USA.}

\begin{abstract}
We present a new method for measuring the $E_G$ statistic that combines two CMB secondaries --- the kinematic Sunyaev–Zel’dovich (kSZ) effect and CMB lensing --- for the first time to probe gravity on linear scales. The $E_G$ statistic is a discriminating tool for modified gravity theories, which leave imprints in lensing observables and peculiar velocities. 
Existing $E_G$ measurements rely on redshift space distortions (RSD) to infer the velocity field. Here, we employ kSZ velocity-reconstruction instead of RSD, a complementary technique that constrains the largest-scale modes better than the galaxy survey it uses. We construct a novel $\hatVG$ estimator that involves a ratio between cross-correlations of a galaxy sample with a CMB convergence map and that with a 3D kSZ-reconstructed velocity field. We forecast for current and upcoming CMB maps from the Atacama Cosmology Telescope (ACT) and the Simons Observatory (SO), respectively, in combination with three spectroscopic galaxy samples from the Dark Energy Spectroscopic Instrument (DESI). 
We find cumulative detection significances in the range $S/N \sim 20-55$, which can robustly test the scale-independent $E_G$ prediction under general relativity (GR) at different effective redshifts of the galaxy samples ($z\approx 0.73, 1.33, 1.84$). In particular, the SO$\times$DESI LRG measurement would be able to distinguish between GR and certain modified gravity models, including Hu-Sawicki $f(R)$ and Chameleon theories, with high confidence. The proposed $\hatVG$ estimator opens up a new avenue for stress-testing gravity and the $\Lambda$CDM+GR model at the largest observable scales. 

\end{abstract}

\maketitle

\section{\label{sec:intro}Introduction}
The Universe's expansion has been accelerating over the last few billion years of its lifetime, a fact firmly established by multiple independent cosmological observations. These include direct constraints from Type Ia supernovae surveys \cite{SN1998, SN1999}, and estimates of the sound horizon scale from Baryon Acoustic Oscillations (BAO) using galaxy redshift survey data at late times \cite{BOSS2017}. At early times, primary anisotropies in the Cosmic Microwave Background (CMB) \cite{wmap, Planck2018} constrain the energy content of different constituents of the Universe and its geometry, providing critical indirect evidence for the acceleration. The `$\Lambda$CDM’, or standard
model of cosmology is the simplest concordance model that successfully explains these and several other observations. 

Dark energy is a general term referring to an additional energy component with negative pressure that is introduced to drive the observed accelerated expansion. In the standard concordant model, dark energy is modeled as a cosmological constant, $\Lambda$, whose energy density remains constant across all times, and dominates the energy budget today ($\Omega_{\Lambda}\approx 0.7$ \cite{Planck2018}). However, current theories of fundamental physics predict a vacuum energy density that is $\sim 120$ orders of magnitude apart from the small $\Lambda$ value measured. As a solution to this `cosmological constant problem', a plethora of \emph{dynamical} dark energy models (such as quintessence e.g.\cite{quinte_1988}) invoking new scalar field(s) have been proposed \cite{Amendola_DE, DEreview2012}. Recent BAO measurements from DR2 of the Dark Energy Spectroscopic Instrument (DESI) \cite{DESI_dr2_bao} found evidence preferring a time-evolving dark energy component over the $\Lambda$CDM model, while combining their constraints with CMB data from the Atacama Cosmology Telescope's DR6 suggests a milder hint for this preference \cite{ACT_dr6_cosmo_2025}.

While cosmologists are gearing up to test the standard $\Lambda$CDM model and dynamical dark energy models more robustly with upcoming datasets, it is timely also to study an alternative explanation of the observed acceleration --- modified gravity. Instead of introducing a new dark energy component, this scenario considers the possibility that general relativity (GR) may not be an accurate theory of gravity on large, cosmological scales. Different modified gravity (MG) models modify GR in ways that cause an apparent accelerated cosmic expansion. They typically involve a screening mechanism (e.g., Chameleon mechanism \cite{Khoury_prd_cham}) that allows GR to be recovered on Solar system scales, where it is on firm footing based on many empirical tests \cite{fR_solarsys_2006, GR_review_2014}. Several MG models today are tightly constrained \cite{GW_2017_MG, JLSchmidt_2016}, especially at nonlinear scales (e.g., the Hu-Sawicki model \cite{HuSawfR2007} of $f(R)$ gravity \cite{Carroll2004}), or ruled out empirically to a large extent (e.g., the flat DGP model \cite{DGP} and its tests \cite{Song2007}). However, it is crucial to robustly test GR against possible MG theories across cosmic redshifts and in various regimes, in particular at linear, cosmological scales. 

To discriminate between dark energy and modified gravity as two possible explanations of the observed accelerated expansion, we need to go beyond `geometrical' probes that only measure the background expansion history \cite{JLSchmidt_2016, Planck15_MG}, since the space of MG models can easily mimic predictions of quintessence dark energy models in this aspect \cite{Lombriser_2016}. Thus, distinguishing tests of gravity also probe the predicted growth of large-scale structure (LSS), which differs between GR and MG models. At subhorizon linear scales, parametrized tests of gravity target modifications to the Poisson equation and the relationship between the temporal and spatial scalar potentials ($\phi$ and $\psi$); these induce scale-dependent signatures in lensing observables and/or in the linear growth rate ($f$) of LSS for several MG models \cite{BZ2008, Hojjati2011, Silvestri2013, Baker2014, HDore2020}.

The so-called `$E_G$' statistic \cite{Zhang2007} is a promising diagnostic test of modified gravity theories at subhorizon linear scales. It consists of a dimensionless ratio between $\potdiff$ and the peculiar velocity field; while lensing observables are sensitive to the integrated $\potdiff$, peculiar velocities are proportional to $f$. In practice, estimators of the $E_G$ statistic measure the cross-correlation of a galaxy sample with a galaxy-lensing \cite{Zhang2007, Reyes2010} or CMB lensing convergence map \cite{Pullen+2015_forecasts}, and take its ratio with respect to the galaxy-velocity cross-power spectrum using the same galaxy field. The concordant $\Lambda$CDM+GR model's prediction of the $E_G$ statistic is only a function of $\Omega_m(z)$, the fractional energy density in matter at the redshift $z$, and is importantly scale-independent. Thus, measurements of the $E_G$ statistic can robustly test \emph{scale-dependent} predictions of MG models at linear scales, while being largely insensitive to the galaxy bias and the amplitude of matter fluctuations \cite{Zhang2007, Amon+2018}. 

Extending the initially proposed estimator of the $E_G$ statistic \cite{Zhang2007}, the first measurement of this quantity was done by analyzing galaxy clustering and galaxy-galaxy lensing using Sloan Digital Sky Survey (SDSS) data \cite{Reyes2010}. Several following measurements of the $E_G$ statistic have used data from various weak galaxy-lensing and spectroscopic galaxy surveys, robustly testing GR at different effective redshifts corresponding to the galaxy sample used (e.g. \cite{Amon+2018, Singh2019, Blake2020, Rauhut_2025}). An estimator utilizing the CMB lensing convergence map instead of galaxy lensing was first introduced in \cite{Pullen+2015_forecasts}; this technique has the benefit of being robust to systematic effects such as intrinsic alignments that are otherwise present in galaxy-lensing-based measurements. Subsequent $E_G$ measurements were made using spectroscopic galaxies and CMB lensing maps from \textit{Planck} \cite{Singh_2020, LWenzl+2024_PlanckBOSS} and most recently from ACT DR6 \cite{LWenzl+2025_ACTBOSS}.

A key point to note is that all estimators and measurements of the $E_G$ statistic to date extract the required peculiar velocity information using redshift space distortions (RSD) \cite{Kaiser_1987} in the 3D galaxy field. In this work, we propose a new `$\hatVG$' estimator of the $E_G$ statistic that combines CMB lensing with another CMB secondary --- the kinematic Sunyaev-Zeldovich (kSZ) effect \cite{SZ1972, SZ1980}. As CMB photons scatter off moving free electrons in the intervening LSS, their Doppler boosting induces a secondary (kSZ) anisotropy in the CMB that is proportional to the integrated electron momentum along the line-of-sight (LOS). The $\hatVG$ estimator involves a ratio between the angular CMB lensing-galaxy power spectrum and an appropriately projected 3D cross-power between the same galaxies and kSZ-reconstructed velocities. This is the first estimator of the $E_G$ statistic that uses the kSZ effect instead of RSD to access peculiar velocities.

kSZ velocity-reconstruction \cite{Deutsch2018, Smith2018} and RSD are complementary approaches for extracting velocity information from galaxy redshift surveys, with different associated systematics (e.g.,\cite{Lamman_2023, giri2020}). However, kSZ tomography has the distinct advantage of constraining the largest-scale modes with lesser noise than the galaxy survey it uses \cite{Smith2018, Munch}. With the arrival of high-resolution CMB maps from the South Pole Telescope \cite{SPT2011}, ACT \cite{ACT2016}, and soon from the Simons Observatory (SO) \cite{SO, ASO_2025}, we can harness the power of kSZ measurements to estimate the velocity field with unprecedented precision. Thus, the proposed $\hatVG$ estimator combines kSZ and CMB lensing with contemporary galaxy data to robustly test GR and probe gravity at the largest possible scales in a new way. 

This paper is organized as follows: Section \ref{sec:Theory} describes the theoretical background of the $E_G$ statistic and its expectation value predicted by GR and a few representative MG models. We carefully construct and derive the novel $\hatVG$ estimator in Section \ref{est}, by combining CMB lensing-galaxy and kSZ derived observables to measure the $E_G$ statistic at a fixed effective redshift. In Section \ref{fore}, we present forecasts for measuring the $E_G$ statistic with the proposed estimator using CMB data from ACT and SO, and spectroscopic galaxy samples from DESI. We present the detection significance of these forecasted measurements for robustly testing the scale-independent $E_G$ prediction from GR at various redshifts. We also discuss the statistical power of such measurements with SO data for distinguishing between GR and certain MG models. We discuss our results in the context of future work and conclude in Section \ref{sec:conclusion}. Throughout this work, unless stated otherwise, the assumed fiducial cosmological model is the concordant $\Lambda$CDM model with best-fit parameters from \textit{Planck} CMB anisotropies \cite{Planck2018}.

\section{Theory}
\label{sec:Theory}
\subsection{The $E_G$ Statistic}\label{sec_EGstat} 
Under General Relativity (GR), a statistically isotropic and homogeneous Universe is described by the Friedmann-Lemaitre-Robertson-Walker (FLRW) metric. For a flat Universe, structure formation and the evolution of perturbations against this smooth background can be derived at large scales (e.g.\,\cite {MaB1995_PT, ModCosmo_Dodelson_2003}) by considering the perturbed FLRW metric: $\mathrm{d}s^{2} = (1+2\psi)\mathrm{d}t^{2} -a^{2}(t)(1+2\phi)\mathrm{d}\textbf{x}^{2}$, where $a$ is the scale factor at a given time, and the two potential fields $\psi$ and $\phi$ denote the time and spatial components respectively.  
The weak lensing convergence $\kappa$ of the CMB is sensitive to $\potdiff$ integrated along the line-of-sight (LOS) (e.g.\,\cite{Lewis2006}):
\begin{equation}
\kappa = \frac{1}{2}\int\mathrm{d}\chi \, \,\chi \left( 1 - \frac{\chi}{\chi_{\mathrm{CMB}}} \right)\,\potdiff,
  \label{kappa}  
\end{equation}
where $\chi(z) \equiv \int_{0}^{z} \mathrm{d}z \,\left(c/H(z)\right)$ is the comoving distance to redshift $z$, $H(z)$ is the Hubble parameter, and $\chi_{\mathrm{CMB}}$ is the comoving radial distance to the surface of last scattering. The equation for $\kappa$ above also holds under any modified gravity model in which photons follow null geodesics \cite{Zhang2007}. In this work, we interchangeably use the scale factor $a \equiv 1/(1+z)$ and the redshift $z$ to refer to the redshift-dependence of any quantity. 

On the other hand, in the linear regime, matter conservation gives the following equation in harmonic space:
\begin{equation}\label{lin-vel}
   \mathbf{v}(\mathbf{k}, z) = i\frac{f(z)aH(z)}{c\,k} \,\delta_m(\mathbf{k},z) \, \hatbf{k}.
\end{equation}
relating peculiar velocities (taken to be dimensionless throughout) and the fractional matter overdensity field  $\delta_m \equiv (\delta\rho_m/\rho_m)$, where $\rho_m$ is the background matter density. Here, $f(z)\equiv(\mathrm{d\,ln}\,D/\mathrm{d\,ln}\,a)$ is the linear growth rate at the redshift $z$ is the scale factor.  Observationally, one way to infer the peculiar velocity field is using redshift space distortions (RSD) in the 3D clustering of galaxies; they are sensitive to the divergence of the comoving velocity field, $\theta \equiv (c\,\nabla\cdot\textbf{v}/aH(z)) = -f(z)\delta_m$.

The $E_G$ statistic \cite{Zhang2007} is a dimensionless quantity constructed to test gravity at subhorizon linear scales, by combining lensing and peculiar velocity information from cosmological observables \cite{JLSchmidt_2016}. First introduced in \cite{Zhang2007}, the $E_G$ statistic is defined in Fourier space as follows: 
\begin{equation}\label{egstat}
    E_G(k, z) \equiv \frac{c^2\,[\nabla^2 (\psi - \phi)(k,z)]}{3H_0^2 (1 + z)\,\theta(k,z)}. 
\end{equation}
In practice, the $E_G$ statistic has been measured at linear scales using estimators \cite{Pullen+2015_forecasts, LWenzl+2024_PlanckBOSS, DG} that take a ratio of the cross-correlation between a galaxy sample and a lensing convergence map with a cross-correlation between the same galaxy sample and a velocity field estimated from RSD. Although galaxies are biased tracers of the underlying matter distribution, taking such a ratio ensures that the linear galaxy bias cancels out, while the predicted effects of modified gravity on matter are tested robustly. 

In this paper (as detailed in Section \ref{estimator_deriv}), we extend this idea to construct a new (`$\hatVG$') estimator for measuring the $E_G$ statistic by using kSZ velocity-reconstruction instead of RSD to extract the velocity field. Using Eq.~\eqref{lin-vel}, we can rewrite $E_G$ statistic in terms of the magnitude of the peculiar velocity field as:  
\begin{equation}\label{vgstat} 
    E_G(k, z) = \frac{c\,H(z)\,[\nabla^2 (\psi - \phi)(k,z)]}{3H_0^2 (1 + z)^2 \, k \,v(k,z)}. 
\end{equation}

\subsection{Predictions of the $E_G$ Statistic: in Modified Gravity Models} \label{predic} 
In GR, in the absence of anisotropic stress, the temporal ($\psi$) and spatial ($\phi$) components of potential perturbations are equal up to a sign difference (convention). Applying the Einstein equations in GR to the perturbed FLRW metric yields the Poisson equation governing the dynamics of non-relativistic matter, which relates the potential ($\psi$) with the distribution of matter. Thus, 
at late times in the Universe (i.e. $z\lesssim10$), these relationships are given by GR in Fourier space as \cite{MaB1995_PT, ModCosmo_Dodelson_2003}: 
\begin{align}\label{GR}
    \nabla^{2}\psi = -k^2 \psi & = 4 \pi G a^2 \rho_{m}(a)  \delta_m, \, \nonumber\\
    \phi & = - \psi, 
\end{align}
where $\rho_{m}(a)$ is the background matter density as a function of the scale factor $a$, and $G$ is the Newtonian gravitational constant.  

Now, for several classes of modified gravity (MG) models, the above equations get modified in a way that can be parameterized as follows at sub-horizon scales \cite{Hojjati2011, Silvestri2013, Baker2014, HDore2020}: 
\begin{align}\label{mugam}
\nabla^{2}\psi = -k^2 \psi & = 4 \pi G a^2 \mu(k, a)  \rho_{m}(a)  \delta_m, \, \nonumber \\
    \phi & = - \gamma(k, a)  \psi.
\end{align}
In this `$\mu-\gamma$ parametrization' of modified gravity models, the two parameters can be functions of time and scale in general. Together, these two parameters phenomenologically determine the impact of deviations from GR on cosmological observables for a given modified gravity model, assuming the quasi-static approximation (which is valid at sub-horizon scales) \cite{JLSchmidt_2016, Zhou2011}. While $\mu$ quantifies the effective strength of gravity (and is equivalent to the related parameter $G_{\mathrm{eff}}$ \cite{HDore2020}), $\gamma$ is the gravitational slip that quantifies the difference between the two potentials. 

It is also common to use the equivalent `$\mu-\Sigma$' reparametrization \cite{Baker2014}, where the function
\begin{equation}\label{sigma}
\Sigma(k, a) \equiv \frac{1}{2}\mu(k, a)(1+\gamma(k, a))
\end{equation}
is used instead of $\gamma(k, a)$. Note that in the special case of GR, $\mu = \gamma = \Sigma = 1$, and we recover the usual equations (Eq.\eqref{GR}). For a given modified gravity scenario, the parameter $\Sigma$ quantifies the overall modification to $\potdiff$, since this is the combination of potentials that lensing observables are sensitive to (Eq.\eqref{kappa}). From Eqs.\eqref{mugam}, it follows that: 
\begin{align}\label{lens-pot}
    \nabla^2 (\psi - \phi) = k^2 (\phi-  \psi)
    & = 3H_0^2\,\Omega_{m, 0}(1+z)\Sigma(k, a)\delta_m.
\end{align}
We have substituted for the background matter density above, where: $\rho_m(z) = \Omega_{m,0}  \frac{3H_0^2}{8\pi G}(1+z)^{3}$, where $\Omega_{m,0} \equiv \rho_{m,0}/\rho_{cr}$ is the fractional energy density in the form of matter today, relative to the critical density $\rho_{cr} \equiv\frac{3H_0^2}{8\pi G}$. 

Following \cite{Pullen+2015_forecasts}, we now derive the expectation value of the $E_G$ statistic for a general MG model that can be characterized using the above $\mu-\Sigma$ parameterization. Starting from the definition in Eq.\eqref{vgstat}, using Eqs.\eqref{lens-pot} and Eq.\eqref{lin-vel}, we obtain the expectation value \cite{Pullen+2015_forecasts}:
\begin{align}\label{EG_mg}
    E_G^{\mg}(k ,z) 
    = \frac{\Omega_{m, 0}  \Sigma(k, z) }{ f(k,z) }.
\end{align}

For a given MG model, the expectation value of the $E_G$ statistic isolates two potentially scale-dependent signatures: the impact of deviation from GR on lensing observables via $\Sigma(k, z)$, and the modification to the Poisson equation via the linear growth rate $f(k, z)$ that is directly dependent on the parameter $\mu(k, z)$ \cite{BP2012}:
\begin{equation}\label{f_ODE}
   a \frac{\mathrm{d}f}{\mathrm{d}a} + f^2 + \left(\frac{1}{2} + \frac{3}{2}(1 - \Omega_{m}(a))\right)f = \frac{3}{2}\Omega_{m}(a)\mu(k, a).
\end{equation}

In the special case of GR (with $\Sigma= \mu= 1$), the predicted value of the $E_G$ statistic is then given by \cite{Zhang2007}:
\begin{align}\label{EG_gr}
    E_G^{\gr}(z) = \frac{\Omega_{m, 0}  }{ f(z) }. 
\end{align}
Assuming GR as the theory of gravity at cosmological scales, and within the concordant standard model of cosmology $\Lambda$CDM, the expected value of the $E_G$ statistic is scale-independent. 
It is obtained by computing the linear growth rate, which is well-approximated as $f(z)\approx \Omega_{m}(z)^{0.55}$ \cite{Linder2005}. Here, $\Omega_{m}(z)=\Omega_{m,0}(1+z)^3\left(H_0/H(z)\right)^2$ is the fractional energy density in matter as a function of redshift. Given a background expansion history that is fixed by external `geometric' measurements, the predicted value of $E_G(z)$ and its associated uncertainty under $\Lambda$CDM+GR can be derived directly from a constraint on $\Omega_{m,0}$ \cite{Pullen+2015_forecasts, LWenzl+2024_PlanckBOSS}. Note that $\Sigma= \mu= 1$ continues to hold in the presence of evolving dark energy modeled by quintessence (e.g.,\cite{quinte_1988}) when GR is the assumed theory of gravity \cite{JLSchmidt_2016}.  

In this work, we consider two particular classes of Modified Gravity (MG) models:
\paragraph{\textbf{Hu-Sawicki $f(R)$ model}} \cite{HuSawfR2007}: $f(R)$ gravity (e.g. \cite{Carroll2004, fR_review}) is one of the most widely studied modifications to Einstein's GR. In this theory, the Einstein-Hilbert action of GR is modified to include an additional term:
\begin{equation}
    S = \int \mathrm{d^4x} \, \sqrt{-g}\,\frac{(R+f(R))}{16\pi G} \, + S_{\mathrm{matter}}, 
\end{equation}
where $f(R)$ is a suitable function of the Ricci scalar $R$, and $g$ is the determinant of the metric tensor. Through a conformal transformation, $f(R)$ gravity can be cast as a scalar-tensor theory of modified gravity \cite{JLSchmidt_2016}. In this context, unlike some earlier models, the Hu-Sawicki $f(R)$ model \cite{HuSawfR2007} is notable for being compatible with precision tests of gravity locally \cite{fR_solarsys_2006}, while also explaining the observed cosmic acceleration. This particular model passes Solar System tests of gravity due to the Chameleon screening mechanism \cite{KhouryW2004_cham}, which activates in regions of high density to hide the effects of this modification locally.     

Assuming a broken power law for $f(R)$ in the Hu-Sawicki model \cite{HuSawfR2007} and imposing that it matches the background expansion history of the standard $\Lambda$CDM model leads to a complete description of the model using two free parameters, $f_{R0}$ and $n$. Phenomenologically, they determine the observational impact of this modification, which is given by the $\mu-\gamma$ parametrization (Eq.\eqref{mugam}) as \cite{Zhao_2009}:
\begin{align}
    \mu^{\mathrm{HS}}(k,a) \equiv \frac{1 + \frac{4}{3}\frac{k^2}{a^2m^2(a)}}{1 + \frac{k^2}{a^2m^2(a)}},\\
    \gamma^{\mathrm{HS}}(k,a) \equiv \frac{1 + \frac{2}{3}\frac{k^2}{a^2m^2(a)}}{1 + \frac{4}{3}\frac{k^2}{a^2m^2(a)}},
\end{align}
where
\begin{align}
    m^2(a)\equiv\left(\frac{H_0}{c}\right)^2\frac{1}{(n+1)|f_{R0}|} \frac{(\Omega_{m,0}a^{-3} + 4(1- \Omega_{m,0}))^{n+2}}{(\Omega_{m,0} + 4(1- \Omega_{m,0}))^{n+1}}
\end{align}
Following \cite{WenzlRoman_2021}, we only focus on the class of models with $n = 1$ in this work; GR is recovered when $f_{R0}\rightarrow0$. Note that with the above forms of $\mu$ and $\gamma$, $\Sigma(k,a) \equiv 1$, and there are thus no modifications to lensing observables.  The Hu-Sawicki $f(R)$ model does, however, predict a scale-dependent deviation from GR for the $E_G$ statistic (Eq.\eqref{EG_mg}), due to the induced scale-dependence in the linear growth rate $f$ driven by $\mu^{\mathrm{HS}}(k,a)$. 

\paragraph{\textbf{Chameleon-type scalar-tensor theories:}} 
In addition to the $f(R)$ gravity models explained above, there are several other models of scalar-tensor theories that belong to the class termed `Chameleon gravity' \cite{KhouryW2004_cham}. In this scenario, a scalar field $\phi$ is non-minimally coupled to matter fields, thus mediating a new `fifth' force exerted on matter particles. The key feature of a Chameleon scalar field is that its effective mass ($m_\phi$) increases in regions having a high local matter density, which \textit{screens} the additional force and thus evades Earth/Solar system tests of gravity. The effective potential of $\phi$ consists of a correction often parametrized as a Yukawa-type potential $\propto e^{-m_{\phi}r}$. In low-density environments, the effective mass of the Chameleon field $m_\phi \sim \mathcal{O}(H_0)$ is low, which thus mediates a long-range force that can potentially drive cosmic acceleration \cite{KhouryW2004_cham, Khoury_prd_cham}.  

We are interested in those Chameleon gravity models that can reproduce the $\Lambda$CDM background expansion history, which is assumed to be fixed. The growth of structure is also modified by these models, which is well-described phenomenologically by the `BZ parametrization' \cite{BZ2008}, and can be written for scalar-tensor theories as \cite{Zhao_2009}:   
\begin{align}\label{cham}
    \mu^{\mathrm{Ch}}(k,a) &\equiv \frac{1 + \beta_1\lambda_1^2k^2a^s}{1 + \lambda_1^2k^2a^s} \\ 
    \gamma^{\mathrm{Ch}}(k,a) &\equiv \frac{1 + \beta_2\lambda_2^2k^2a^s}{1 + \lambda_2^2k^2a^s},
\end{align}
where $\beta_2 = (2/\beta_1) - 1$ and $\lambda_2^2 = \beta_1 \lambda_1^2$. Chameleon models can thus be characterized by a tuple $(\beta_1, \lambda_1^2, s)$ 
The typical range of $\beta_1$ is (0,2), and it represents a dimensionless coupling \cite{Zhao_2009, Hojjati2011}. $\lambda_1$ has dimensions of length 
and can be equivalently expressed in terms of the parameter $0 \leq B_0 \equiv 2\lambda_1^2H_0^2/c^2 \leq 1$ \cite{Hojjati2011, Pullen+2015_forecasts}. The time evolution of the mass of the scalar field determines the parameter $s$; its typical range is $(0,4]$.
Note that with the above parametrization, the combination $\Sigma=1$; so while lensing observables are not modified, the $E_G$ statistic's prediction in Chameleon gravity differs from GR due to $f(k,a)$ (Eq.\eqref{f_ODE})

Aside from the two classes of MG models considered in this work, the $E_G$ statistic has also been explored as a tool to distinguish between GR and other possible MG theories at cosmological scales, including certain TeVeS models \cite{Zhang2007, Reyes2010}, DGP models \cite{Zhang2007}, and other $f(R)$ gravity models \cite{Zhang2007, Pullen+2015_forecasts}.

\section{Novel estimator for the $E_G$ statistic: using kSZ Tomography}\label{est} 
In this section, we construct and propose a new estimator for measuring the $E_G$ statistic (Eq.\eqref{vgstat}) observationally, to test gravity on linear scales. While we follow \cite{Pullen+2015_forecasts} and use CMB lensing to access modifications to $\potdiff$, the novelty of our approach lies in using kSZ velocity-reconstruction instead of RSD to extract peculiar velocity information in a complementary way.    

\subsection{kSZ Velocity-Reconstruction}\label{ksz}
As CMB photons traverse through the Large-Scale Structure (LSS) and Compton-scatter off free electrons moving with a bulk velocity, a secondary anisotropy proportional to the electron momentum is induced in the CMB, known as the kinematic Sunyaev-Zel’dovich (kSZ) effect \cite{SZ1972, SZ1980, Ostriker1986}. In a direction $\hatbf{n}$ on the sky, the kSZ anisotropy is expressed as:
\begin{equation}
    \Theta^{\mathrm{kSZ}}\left( \hatbf{n}\right) \equiv \frac{\Delta T^{\mathrm{kSZ}}\left( \hatbf{n}\right)}{T_{\mathrm{CMB}}} 
    = \int \mathrm{d}\chi \, K(z) \, (1+\delta_{e}(\mathbf{x})) v_{r,e}(\mathbf{x}),
\end{equation}
where $\delta_e$ is the electron overdensity, and $v_{r,e}$ is the electron velocity field along the LOS. The kSZ prefactor here is given as: $K(z) \equiv -\sigma_{\mathrm{T}}n_{e,0}x_e(z)(1+z)^2e^{-\tau(z)}$, where $\sigma_{\mathrm{T}}$ is the Thomson cross-section, $n_{e,0}$ is the number electron density today, $x_e$ is the ionization fraction, and $\tau(z)$ is the optical depth at redshift $z$. 

The kSZ signal sourced at late-times can be detected in CMB maps by combining with some LSS data through various estimators. Among these, `kSZ tomography' refers to those techniques (e.g. \cite{Ferreira1999, Ho2009, Shao, Alonso2016, Deutsch2018}) that are sensitive to an underlying signal of the form $\langle Tgg\rangle$, i.e., a bispectrum between one power of the CMB temperature field (`$T$') and two powers of a galaxy field (`$g$') \cite{Smith2018}\footnote{While the techniques of pairwise-velocity, velocity-weighted stacking, and velocity-reconstruction, among others, were found to be mathematically equivalent to $\langle Tgg\rangle$ in \cite{Smith2018}, not all kSZ estimators are of this form. In particular, `projected-fields' kSZ estimators that are applicable to projected-fields of various LSS tracers are either of the form $\langle T^{2}\times g\rangle$ \cite{Dore2004, D05, Bolliet2022, P23} or $\langle TTg\rangle$ \cite{P25_bis}.}. 

The tomographic approach of kSZ `velocity reconstruction' \cite{Deutsch2018} is well-suited for cosmological applications (e.g.\,\cite{Smith2018, Munch}). It consists of a quadratic estimator of the \textit{large-scale} radial velocity field $v_r$, which crosses one power of the CMB temperature map with one power of a galaxy field at small scales ($k_S\sim1$ Mpc$^{-1}$). When this reconstructed velocity field $\vrad$ is cross-correlated with a 3D galaxy field on large scales ($k\lesssim$ 0.1 Mpc$^{-1}$), the result is equivalent to an optimal estimator of the $\langle Tgg\rangle$ bispectrum for squeezed triangles (having one long-wavelength `$g$' side, and two short-wavelength sides). Here, we restrict our analysis to 3D velocity-reconstruction with spectroscopic galaxy samples only (unlike some other kSZ analyses \cite{BlochJohnson2024, FMcCarthy2024}); we refer the reader to \cite{Deutsch2018, Smith2018} for further details of this approach.   

Following the framework of \cite{Smith2018, giri2020}, we assume a simplified geometry of a periodic box where the Universe is ``snapshotted" at a particular time $t_*$ with a redshift $z_*$; when considering a specific galaxy survey, the true geometry is approximated by matching the box volume $V$ and $z_*$ to the comoving volume and effective redshift ($\zeff^{gg}$) of the galaxy sample, respectively. The first 3D measurement of kSZ velocity reconstruction \cite{Lague2024} was performed in ACT data using SDSS galaxies by following this framework. 
Another measurement of 3D velocity reconstruction was made recently using galaxy data from DESI Legacy Imaging Surveys, with an improved pipeline that accounts for lightcone evolution along with a sky cut \cite{Hotinli2025}. However, for the purpose of our forecasts, we continue to assume the simplified `snapshot' geometry. 

Working in harmonic space, the Fourier transforms of the CMB temperature map (in 2D) and the galaxy overdensity field (in 3D) are denoted as $T(\boldsymbol{\ell})$ and $\delta_g(\mathbf{k})$, respectively. Here, we consider kSZ velocity reconstruction at large scales only ($k\lesssim$ 0.1 Mpc$^{-1}$); in this regime, linear theory is accurate, and velocities of electrons match the underlying peculiar velocity field, which is curl-free. Thus, from the continuity equation (Eq.\eqref{lin-vel}), the reconstructed radial velocity $\vrad$ in 3D can be converted to a reconstruction of the total magnitude of the velocity field $v(\mathbf{k})$, or the matter density field $\delta_m(\mathbf{k})$ \cite{Smith2018}:  
\begin{equation}
 v_r(\mathbf{k}) = i\frac{k_r}{k}v(\mathbf{k}) = \mathrm{cos}(\vartheta)\frac{faH}{c\,k} \,\delta_m(\mathbf{k}),   
\end{equation}
where $k_r$ is the radial component of $\mathbf{k}$ (i.e. when $\mathbf{k}$ is projected along the LOS), and $\mathrm{cos}(\vartheta)\equiv k_r/k$. Assuming a statistically isotropic Universe, the anisotropic two-point cross-correlation between the reconstructed radial $\vrad$ and a 3D galaxy field $\delta_g$ can be written as:
\begin{equation}
    \langle\delta_g(\mathbf{k'})*\vrad(\mathbf{k})\rangle = i\frac{k_r}{k}P_{vg}(k)(2\pi)^3\delta^3_{D}(\mathbf{k}-\mathbf{k'}),
\end{equation}
where $P_{vg}(k)$ is the 3D cross-power spectrum between $\delta_g(\mathbf{k})$ and the magnitude of the velocity field $v(\mathbf{k})$. The corresponding error bars on the observable $P_{vg}(k)$ (measured at $\zeff^{gg}$) are given by (see Section V.C in \cite{Smith2018}):
\begin{align}\label{var Pvg}
    \sigma(P_{vg}(k)) & = \left( {V}\int_{k_{\min}}^{k_{\max}}\int_{-1}^{1}  \frac{1}{4\pi^2}\frac{k^2  \mathrm{d}k \ \mathrm{d}\mur}{P_{g g}^{\mathrm{tot}}(k)  N^{\mathrm{rec}}_{vv}(k,\vartheta)}\right)^{-1/2} \nonumber\\
    & = \left( \frac{2V}{3}\int_{k_{\min}}^{k_{\max}}  \frac{k^2\mathrm{d}k }{4\pi^2} \frac{1}{P_{g g}^{\mathrm{tot}}(k)  N_{v_r}(k)} \right)^{-1/2},
\end{align}
where $P_{vg}(k)$ is measured in the 3D $k$-bin=($k_{\min}$, $k_{\max}$), and $N^{\mathrm{rec}}_{vv}(k,\vartheta) = (\mur)^{-2}N_{v_r}(k)$ is the noise power spectrum for the total magnitude of the velocity field in that $k$-bin. $N_{v_r}(k)$ denotes the noise power spectrum of the kSZ-reconstructed radial velocity field, which tends to a $k$-independent constant in the assumed `squeezed limit' ($k\lesssim 0.1$/Mpc, $k_S\sim1$-2.5/Mpc) of the $\langle Tgg\rangle$ bispectrum \cite{Smith2018, giri2020}:
\begin{equation}\label{Nvr}
    N_{v_{r}}(k) = \frac{\chi^2_{\mathrm{eff}}}{K^2_{\mathrm{eff}}}\left[\int
    \frac{k_s\,\mathrm{d}k_s}{2\pi}\left(\frac{P_{g e}^2(k_s)}{P_{gg}^{\mathrm{tot}}(k_s)C_\ell^{TT, \mathrm{tot}}}\right)_{\ell\approx k_s\chi_{\mathrm{eff}}} \right]^{-1},
\end{equation}
where $\chieff \equiv \chi(\zeff)$ and $\Keff \equiv K(\zeff)$ are the comoving radial distance and the kSZ prefactor evaluated at $\zeff^{gg}$. $C_\ell^{TT, \mathrm{tot}}$ is the total power spectrum of the component-seperated CMB map, which includes the noise power spectrum for the specific CMB experiment that is considered. $P_{gg}^{\mathrm{tot}}(k_S)$ is the total 3D galaxy power spectrum, which includes a shot noise term specific to the galaxy sample. 

Note that the optimal weights used while computing the minimum-variance quadratic estimator for $\vrad$ \cite{Deutsch2018, Smith2018} and its associated velocity reconstruction noise given above (Eq.\eqref{Nvr}) both depend on the assumed form of the galaxy-electron power spectrum $P_{ge}$ at small scales, which is dictated by the distribution of electrons (baryons). If the choice of model for the baryon density profile differs from the true profile due to uncertainties in baryonic feedback, this introduces a bias $b_v$ in the reconstructed velocities that is constant on large enough scales ($k\lesssim$ 0.1/Mpc): $\langle\hat{v}_r(\kvec)\rangle = b_v\,\vrad(\kvec)$ \cite{Smith2018}; if the true electron density profile matches the assumed fiducial one, then $b_v = 1$. 

Thus, \textit{large-scale} velocity reconstruction can only be performed by assuming a particular electron density profile at \textit{small scales}. This is a subtlety known as the `optical depth degeneracy', which allows a normalization amplitude to be exchanged between the measured $P_{vg}(k)$ at large scales and $P_{ge}(k_S)$. In practice, using an external measurement of the electron density profile (e.g. with kSZ velocity-weighted stacking \cite{Hadzhiyska2024, BRGuechella2025} or using Fast Radio Bursts (FRBs)\cite{Madhav_frb_2019}) as the fiducial model can limit 
this degeneracy. This would provide direct kSZ measurements of the velocity field that have lower noise than the galaxy survey (and RSD) on the largest scales \cite{Smith2018, Munch}, through sample variance cancellation \cite{Seljak2009}. Additionally, note that even in the presence of a constant velocity bias $b_v \neq 1$, the scale-dependence of the measured $E_G$ statistic remains unaffected.   

\subsection{CMB Lensing-Galaxy Angular Cross-power Spectrum} 
\label{sub:cmb_lensing}
As explained above, we extract the velocity-galaxy power spectrum ($P_{vg}$) using the kSZ effect to probe modifications to the growth rate of structure (Eqs.\eqref{vgstat}, \eqref{f_ODE}). For the other piece of the puzzle, we follow \cite{Pullen+2015_forecasts, LWenzl+2024_PlanckBOSS,LWenzl+2025_ACTBOSS} and use CMB lensing to test the imprints of modified gravity on $\potdiff$ (Eqs.\eqref{vgstat},\eqref{EG_mg}). 

Throughout this work, we consider galaxy clustering as a biased tracer of the underlying matter distribution to obtain the cross-correlations needed to measure the $E_G$ statistic at large scales ($k\lesssim0.1/$Mpc). In this regime, the galaxy overdensity field $\delta_g = b_g\delta_m$ traces the underlying matter overdensity, with a linear galaxy bias, $b_g$. Thus, linear power spectra involving the galaxy field ($g$) can be expressed as: $P_{mg} = b_gP_{mm}$, $P_{\potdiff g} = b_gP_{\potdiff m}$, and $P_{gg}=b_g^{2}\,P_{mm}$. 

Now, even though the $E_G$ statistic is defined in 3D, the CMB lensing convergence $\kappa$ is only accessible as a 2D projected map, with contributions from $\potdiff$ integrated (radially) along the LOS (Eq.\eqref{kappa}). Thus, the relevant observable in our context is the CMB Lensing-galaxy angular cross-power spectrum, $\kgCl$, which is a 2D projection on the sky of the underlying 3D cross-power spectrum $P_{\potdiff g}$. We work in the flat-sky limit 
($\ell \geq 40$ throughout this paper), in which $\kgCl$ is described by the Limber approximation \cite{Limber, Limber-ext} as:  
\begin{align}\label{Kg}  
   C_{\ell}^{\kappa g} & = \int \mathrm{d}z\frac{\widehat{W}_{\kappa}(z)W_{g}(z)}{\chi^2(z)} \, \frac{1}{2(1+z)} P_{\potdiff g}\left(k_{\ell}(z), z \right) \nonumber \\ 
   &\stackrel{\mathrm{(GR)}}{=} \int \mathrm{d}z  \frac{W_{\kappa}(z)W_{g}(z)}{\chi^2(z)} P_{mg}\left(k = k_{\ell}(z), z \right), 
\end{align}
where $k_{\ell}(z) \equiv \frac{\ell + 0.5}{\chi(z)}$. The first equation above holds in a general modified gravity scenario (Eq.\eqref{kappa}) and the (generalized) CMB lensing kernel is defined as \cite{Pullen+2015_forecasts, LWenzl+2024_PlanckBOSS,LWenzl+2025_ACTBOSS}: 
\begin{align}
    \widehat{W}_{\kappa}(z) \equiv (1+z)  \chi   \left( 1 - \frac{\chi}{\chi_{\mathrm{CMB}}} \right).
\end{align}

The lower relation in Eq.\eqref{Kg} above holds in the special case of GR as the theory of gravity, where $\potdiff$ is directly related to the matter overdensity field (Eqs.\eqref{GR}). The corresponding CMB lensing kernel under GR is typically defined (with respect to the matter overdensity) as \cite{Lewis2006}: $W_{\kappa}(z) = \frac{3H_0^2}{2c^2}  \Omega_{m, 0}\widehat{W}_{\kappa}(z)$. For angular power spectra involving galaxies (including Eq.\eqref{Kg} above), the window function $W_g(z)$ for a particular galaxy sample characterizes the relative contributions from different $z$-bins that are integrated along the LOS, and is defined as:   
\begin{equation} 
    W_{g}(z) \equiv \left\langle \dfrac{\mathrm{d}N}{\mathrm{d}z} \right\rangle,
\end{equation}
where $\frac{\mathrm{d}N}{\mathrm{d}z}$ denotes the redshift distribution of the number of galaxies, which is normalized above such that it sums up to 1.

The uncertainty (variance) of the observable $\kgCl$ is given analytically as (following \cite{Pullen+2015_forecasts}):
\begin{align}\label{var-Kg}
    \sigma^2(\kgCl) = \frac{\left(\kgCl\right)^2 + (C_{\ell}^{\kappa \kappa} + N_{\ell}^{\kappa \kappa})(C_{\ell}^{gg} + N^{gg})}{(2\ell + 1) f_{\mathrm{sky}}},
\end{align}
where $N_{\ell}^{\kappa \kappa}$ is the noise power spectrum of the CMB convergence map from lensing reconstruction, and $N^{gg}$ is the (2D) shot noise of the galaxy sample. $C_{\ell}^{\kappa \kappa}$ and $C_{\ell}^{gg}$ are the angular auto-power spectra of the CMB convergence map and galaxy field, respectively, and are given by the Limber approximation \cite{Limber, Limber-ext} as: 
\begin{align}\label{Clgg}
    \ggCl = \int \mathrm{d}z  \frac{H(z)}{c}  \frac{W_g^2(z)}{\chi^2(z)}  P_{gg} \left( k = k_{\ell}(z), z \right),
\end{align}
and 
\begin{align}\label{Clkk}
    C_{\ell}^{\kappa \kappa} 
    \stackrel{\mathrm{(GR)}}{=} \int \mathrm{d}z \, \frac{c}{H(z)} \frac{W_{\kappa}^2(z)}{\chi^2(z)} P_{mm}\left(k = k_{\ell}(z), z \right).
\end{align}

\subsection{Effective Redshifts}\label{zeff}
In the subsequent subsection, we will construct an estimator of the $E_G$ statistic by combining $\kgCl$ with a quantity derived from the kSZ-reconstruction observable, $P_{vg}$. Since $\kgCl$ is an angular power spectrum that integrates the underlying 3D power spectrum $P_{mm}$ (in GR) across the redshift range of the galaxy sample, the overall effective redshift of this observable is defined as \cite{Chen2022}: 
\begin{align}
    \zeff^{\kappa g} \equiv \frac{\int \mathrm{d}z \ \chi^{-2} \  \widehat{W}_{\kappa}(z) W_g(z) \ z }{\int \mathrm{d}z \ \chi^{-2} \ \widehat{W}_{\kappa}(z) W_g(z) }.
\end{align}

Meanwhile, as detailed in Section \ref{ksz}, in the simplified geometry framework of kSZ tomography \cite{Smith2018, Lague2024}, the `snapshot' redshift associated with the reconstructed velocity field ($z_*$) is set to be equal to the effective redshift of the spectroscopic galaxy sample used. Since the effective redshift of the (2D) angular galaxy power spectrum matches that of the 3D clustering of the same galaxy sample \cite{Chen2022}, the effective redshift of the kSZ-derived observable is given by: 
\begin{align}
    \zeff^{gg} \equiv \frac{\int \mathrm{d}z \ \chi^{-2} \  W_g^2(z) \  (H(z)/c) \ z }{\int \mathrm{d}z \ \chi^{-2} \ W_g^2(z) \  (H(z)/c) }.
\end{align}

Note that the $\zeff^{gg}$ and $\zeff^{\kappa g}$ of a galaxy sample do not match in general. As pointed out in previous works on the $E_G$ statistic \cite{Pullen+2016_PlanckCMASS, Singh2019, Blake2020}, this mismatch can potentially introduce a bias in the measurement which otherwise needs to be corrected. For our proposed estimator, we follow an approach similar to the one prescribed in \cite{LWenzl+2024_PlanckBOSS,LWenzl+2025_ACTBOSS} to ensure that it is unbiased. Specifically, while evaluating the kSZ-derived observable, we use a reweighted 3D galaxy sample `$\gdag$', which is obtained from the original spectroscopic galaxy sample ($g$) by modifying its window function:  
\begin{align}
   W_{\gdag}(z) =  \left\langle \dfrac{\mathrm{d}N^{\dagger}}{\mathrm{d}z} \right\rangle \equiv \left\langle \dfrac{\mathrm{d}N}{\mathrm{d}z} \right\rangle  \omega(z) = W_{g}(z) \omega(z),
   \label{eqn:Wg_dagger_definition}
\end{align} 
where the multiplicative weights introduced are defined as:
\begin{align}
    \omega(z) \equiv \frac{1}{I}\sqrt{\frac{\widehat{W}_{\kappa}(z)}{W_g(z)}  \frac{c}{H(z)}},
\end{align}
and the corresponding normalization is:
\begin{align}
    I \equiv \int \mathrm{d}z \ \left\langle \dfrac{\mathrm{d}N}{\mathrm{d}z} \right\rangle \sqrt{\frac{\widehat{W}_{\kappa}(z)}{W_g(z)}  \frac{c}{H(z)}}.
\end{align}

The above choice of reweighing the galaxy sample while computing the kSZ-derived observable ensures that the effective redshift of the reconstructed velocity field, $z_{\text{eff}}^{g^{\dagger}g^{\dagger}}$, matches the effective redshift of the lensing observable $z_{\text{eff}}^{\kappa g}$, and that the proposed estimator is unbiased. For the remainder of this paper, we drop the superscripts for brevity of notation and use $\zeff$ to denote $\zeff^{\kappa g} = \zeff^{g^{\dagger}g^{\dagger}}$, i.e., the overall effective redshift of the $E_G$ statistic's estimator.

\subsection{Defining the new $\widehat{V}_G$ Estimator}\label{estimator_deriv}
Starting from the definition in Eq.\eqref{egstat}, an exact estimator in 3D with expectation value equal to the $E_G$ statistic would be \cite{Pullen+2015_forecasts}:
\begin{equation}\label{vg_est_exact} 
    E_G(k, z) = \frac{c\,H(z)\,P_{\nabla^2 (\psi - \phi)g}(k,z)}{3H_0^2 (1 + z)^2 \, k \,P_{vg}(k,z)}. 
\end{equation}
Now, as noted earlier, the 3D power spectrum $P_{\nabla^2 (\psi - \phi)g}(k,z)$ is only observable here as an integrated quantity along the LOS, i.e. the angular cross-power spectrum between the CMB lensing convergence map and the galaxy field, $\kgCl$ (Eqs.\eqref{kappa},\eqref{Kg}). Therefore, for a fair comparison, we construct a corresponding quantity resembling angular power spectra, 
$\vgCl$, by projecting the 3D kSZ observable ${P}_{v\gdag}(\kvec, \zeff)$ (Section \ref{ksz}) as follows:
\begin{align}
    \vgCl\equiv\int\frac{\mathrm{d}z\,\widetilde{W}_{v}(z) W_{\gdag}(z)}{\chi^2(z)}     \frac{(1+z)}{H(z)}k_{\ell}(z)P_{v\gdag}\left(k_{\ell}(z)\right) \label{eqn:Clvg_def} 
\end{align}

To compute this kSZ-derived quantity, we use the same galaxy sample as in $\kgCl$ with an additional reweighing (i.e. $\gdag$; Eq.\eqref{eqn:Wg_dagger_definition}), to match the effective redshifts of the two observables as explained in the previous subsection. Here, we have defined the window-like function $\widetilde{W}_{v}$ corresponding to the velocity field as:
\begin{align}
    \widetilde{W}_{v} \equiv \frac{\widehat{W}_{\kappa}(z)}{\omega(z)}, \label{eqn:Wv_definition}
\end{align}
so that it follows straightforwardly from Eq.~\eqref{eqn:Wg_dagger_definition} that
\begin{equation}
  \widetilde{W}_{v}(z) W_{\gdag}(z) = \widehat{W}_{\kappa}(z) W_g(z).  
\end{equation}
Thus, we have constructed $\vgCl$ from the 3D kSZ observable $P_{v\gdag}$ by using the same redshift-weighting (i.e. product of projection kernels) as the lensing observable $\kgCl$. 
This ensures that we do not induce a scale-dependent multiplicative bias through the LOS projection (as pointed out in \cite{Pullen+2015_forecasts, LWenzl+2024_PlanckBOSS,LWenzl+2025_ACTBOSS}). Thus, we define our proposed estimator $\hatVG$ that combines CMB lensing and kSZ velocity-reconstructed observables at linear scales as:
\begin{align}
    \hatVG(\ell, z_{\mathrm{eff}}) = \left(\frac{2c}{3H_0^2}\right) \frac{C_{\ell}^{\kappa g}}{\vgCl},
\end{align}
whose expectation value is the dimensionless $E_G$ statistic. 

Similar to other estimators of the $E_G$ statistic \cite{Zhang2007, Pullen+2015_forecasts, LWenzl+2024_PlanckBOSS,LWenzl+2025_ACTBOSS}, measurements with the proposed $\hatVG$ estimator are expected to be insensitive to the galaxy bias as well as the amplitude of initial matter fluctuations in the linear regime.  
Furthermore, since we can directly measure the galaxy-velocity cross-power spectrum with kSZ velocity-reconstruction, we make fewer approximations while deriving the $\hatVG$ estimator; in contrast, previous estimators \cite{Reyes2010, Pullen+2015_forecasts, LWenzl+2024_PlanckBOSS} typically measure this quantity by splitting it as a product of the RSD parameter ($\beta = f/b_g$) and $C_{\ell}^{gg}$. The reweighing of galaxies for $\vgCl$ is also defined in a way such that we do not need to assume that $H(z)$ is slow-varying across the redshift ranges considered, especially for deeper galaxy samples. 

Our proposed estimator to measure the $E_G$ statistic is complementary to previous methods that infer the velocity information from galaxy surveys alone (via RSD), and is thus affected by distinct systematic effects. Moreover, for the largest-scale modes, the kSZ-reconstructed velocity field has lower noise as compared to that derived from the galaxy sample itself \cite{Smith2018, Munch}. 

As discussed in Section \ref{ksz}, kSZ tomography allows the overall normalization of $P_{gv}(k)$ to be varied (characterized by a scale-independent $b_v$) due to uncertainty in the electron density profile. An accurate external measurement of $P_{ge}$ at small scales using the same DESI galaxies can ensure a velocity bias $b_v$ close to 1 (e.g. with kSZ velocity-weighted stacking \cite{Hadzhiyska2024, BRGuechella2025}, or by breaking the degeneracy using Fast Radio Bursts \cite{Madhav_frb_2019}). Importantly, even when the fiducial and true $P_{ge}$'s do not match, the presence of a constant velocity bias $b_v \neq 1$ in the squeezed limit would \textit{not} alter the scale-dependence of the measured $E_G$ statistic. Thus, the proposed $\hatVG$ estimator is a robust test of the scale-independent prediction of the $E_G$ statistic under the standard GR$+\Lambda$CDM model.  
\subsection{Covariance Matrix: Analytical Form}
\label{sub:covariance_matrix}
Starting from the definition of the proposed $\hatVG$ estimator (Eq.~\eqref{vgstat}) as a ratio of the lensing observable $\kgCl$ and the kSZ-derived quantity $\vgCl$, we \emph{initially} assume that the two are measured independently and are uncorrelated. The fractional error on the $E_G$ statistic estimated with this method is then given analytically as:
\begin{align}\label{var VG_est} 
    \frac{\sigma^2\,[\hatVG(\ell, \zeff)]}{E_G^2(\ell, \zeff)} = \left( \frac{\sigma(\kgCl)}{\kgCl}\right)^2 + \left( \frac{\sigma(\vgCl)}{\vgCl}\right)^2,
\end{align}
where the expectation value of the $\hatVG$ estimator matches the $E_G$ statistic. We describe the numerical computation of all the required quantities for our experiments of interest in Section \ref{survey}. The total uncertainty in the angular CMB lensing-galaxy power spectrum, $\sigma(\kgCl)$, is analytically given by Eq.~\eqref{var-Kg}.

To estimate the uncertainty on the kSZ-derived quantity, $\vgCl$, we first express the integral in Eqn.~\eqref{eqn:Clvg_def} as a sum over redshift bins (each of width taken to be $\Delta z = 0.1$ here), to obtain:
\begin{align}
    \vgCl \approx (\ell+0.5)  \Delta z  \ \sum_{i} \left[ \alpha_i P_{v\gdag}\left(k_{\ell}(z_i), \zeff \right)\right], \label{eqn:Clvg_approx_sum}
\end{align}
where the overall redshift `weighting' associated with the $i^{\mathrm{th}}$ redshift bin is defined for convenience as:
\begin{align}
     \alpha_i \equiv \frac{\widetilde{W}_{v}(z_i) W_{\gdag}(z_i) (1 + z_i)}{\chi^3(z_i) H(z_i)}.
\end{align}
As explained in Section \ref{ksz}, in the simplified snapshot geometry framework, kSZ velocity-reconstruction measures the 3D $P_{v\gdag}(k, \zeff)$ as a function of $k$ in a 3D box centered at $\zeff^{g^{\dagger}g^{\dagger}} = \zeff^{\kappa g} \equiv \zeff$; the total uncertainty on this kSZ observable is calculated in $k$-bins by substituting the reweighed galaxy sample $\gdag$ in Eq.\eqref{var Pvg}.   

The total variance of the constructed quantity $\vgCl$ is then described analytically as:
\begin{equation}
    \sigma^2\left(\vgCl\right) \approx \left( \ell +0.5\right)^{2}\Delta z^{2}\sum_{i} \alpha^2_i \sigma^2\left(P_{v\gdag}\left(k_{\ell}(z_i)\right) \right) 
\end{equation}
\\
This follows because for a fixed $\ell$, $P_{v\gdag}\left(k_{\ell}(z_i)\right)$ is uncorrelated across separate redshift bins. 

Given the high detection significance forecasted for kSZ velocity-reconstruction with high-resolution current ACT data, and upcoming SO data of even higher sensitivity \cite{Smith2018}, we find that the overall error budget of the $\hatVG$ estimator (Eq.~\eqref{var VG_est}) is dominated by the error on $\kgCl$ (refer to Section \ref{fore} for further discussion with forecasted results)\footnote{The lensing-reconstruction noise also dominates the overall error of previous estimators employing CMB lensing, in combination with RSD instead \cite{Pullen+2015_forecasts, LWenzl+2024_PlanckBOSS}.}. Now, suppose we relax the assumption used in Eq.~\eqref{var VG_est} where $\kgCl$ and $\vgCl$ are uncorrelated quantities, and instead consider the scenario where they are fully correlated (i.e. the absolute value of their correlation coefficient is 1). Then, the relative uncertainty on the $\hatVG$ measurement would be the sum of the absolute values of the relative uncertainties of $\kgCl$ and $\vgCl$, instead of Eq.~\eqref{var VG_est}. In this extreme scenario, we find that the error on $\hatVG$ is slightly higher, and the resulting cumulative SNRs for SO (ACT) measurements reduce by $< 3\%$ ($< 6\%$).   

Also, in practice, the $E_G$ statistic would be measured using the $\hatVG$ estimator with an $\ell$-binning. Even in this binned case, the errors on $\hatVG(\ell)$ in adjacent $\ell$-bins would be slightly correlated since they share some $k$-modes. Thus, to accurately interpret a measurement with the $\hatVG$ estimator in the future, it would be important to estimate the covariance matrix of $\hatVG$ from corresponding simulations to quantify the non-diagonal terms, as done in recent measurements with an existing, separate estimator \cite{LWenzl+2024_PlanckBOSS, LWenzl+2025_ACTBOSS}. Since the overall error budget of $\hatVG$ is dominated by the uncertainty in $\kgCl$, we expect that the analytical expression in Eq.~\eqref{var VG_est} would be a good approximation of the \emph{diagonal} elements of $\hatVG$'s covariance (and as explained above, even if $\kgCl$ and $\vgCl$ are highly correlated, the resulting cumulative SNRs would be lowered by $<6\%$). While the off-diagonal terms are expected to be smaller than the variance, it would be valuable for future works to compute the simulated covariance matrix of $\hatVG$ to validate our approach and to accurately account for off-diagonal terms. 
For the purpose of our forecasts here, we will continue to assume the analytical expression of $\hatVG$'s variance (Eq.~\eqref{var VG_est}) for simplicity.

\section{Forecasts} \label{fore} 
We now present forecasts for measuring the $E_G$ statistic using data in the near future, with the proposed $\hatVG$ estimator, as constructed in the previous section. 

\subsection{Survey Specifications and Numerical Implementation}\label{survey}
Since the proposed $\hatVG$ estimator includes peculiar velocities extracted using the kSZ effect, which dominates in cleaned CMB maps around arcminute scales, we consider two high-resolution CMB experiments. Firstly, we forecast for publicly available current CMB data from ACT-DR6, the latest data release of the Atacama Cosmology Telescope (AdvACT) \cite{ACT2016}. We also forecast for upcoming CMB data from the Large Aperture Telescope of the Simons Observatory (SO) \cite{SO}, which is currently observing from Cerro Toco in Chile. 

To compute the uncertainty of the angular CMB-lensing cross galaxy power spectrum $\kgCl$ (Eq.~\eqref{var-Kg}), we use publicly available baseline curves for $N_{\ell}^{\kappa \kappa}$, the noise power spectrum of the CMB convergence map after lensing reconstruction, for ACT DR6\footnote{\url{https://lambda.gsfc.nasa.gov/product/act/actadv_dr6_lensing_maps_info.html}} \cite{ACTDR6_Qu2024}, and as predicted for SO\footnote{\url{https://github.com/simonsobs/so_noise_models/blob/master/LAT_lensing_noise/lensing_v3_0_0/Apr17_mv_nlkk_deproj0_SENS1_fsky_08000_iterOn.csv}} \cite{SO}. While computing the noise of kSZ-reconstructed velocities (Eq.~\eqref{Nvr}), $C_\ell^{TT, \mathrm{tot}} = (C_\ell^{TT} + C_\ell^{\mathrm{kSZ}} + N_\ell^{TT})$ is the total power spectrum of the component-seperated CMB map, where we use a realistic post-ILC noise curve derived from simulated maps for the noise power spectrum $N_\ell^{TT}$ of SO\footnote{https://github.com/simonsobs/so_noise_models/}. For ACT DR6 (AdvACT), following \cite{Bolliet2022}, we model it as a white noise power spectrum for simplicity:
\begin{equation} \label{ACT}
  N_\ell^{TT} = \Delta_T^2 \, \mathrm{exp} \left(\ell(\ell+1)\frac{\theta_{\mathrm{FWHM}}^{2}}{8\,\mathrm{ln}\,2} \right),
\end{equation}
where the resolution of AdvACT corresponds to $\theta_{\mathrm{FWHM}}$ = 1.5 arcmin, while the noise level is inflated beyond the raw map's noise to account for the impact of component-separation, and is taken to be $\Delta_T = 20 \,\mu K$-arcmin. The (theoretical) kSZ auto-power spectrum $C_\ell^{\mathrm{kSZ}}$ is computed using a template\footnote{\url{https://github.com/nbatta/SILC/blob/master/data/ksz_template_battaglia.csv}} derived from hydrodynamical simulations \cite{Battaglia2010}. For both experiments, we compute the lensed primary CMB power spectrum, $C_\ell^{TT}$, using the code CAMB \cite{camb1, camb2}, while assuming the fiducial $\Lambda$CDM+GR model. 

We also compute the required background quantities (e.g. $\chi(z)$), and the linear matter power spectrum $P_{mm}(k)$ appearing in Eqs.~\eqref{var-Kg},~\eqref{Kg},~\eqref{Clgg},~\eqref{Clkk}, ~\eqref{eqn:Clvg_def}, and ~\eqref{var Pvg} by assuming the same fiducial cosmological model in CAMB. In this work, we consider separate measurements using three different spectroscopic galaxy samples from the Dark Energy Spectroscopic Instrument (DESI) \cite{DESI_expt}, which is currently midway through making observations in its five-year-long main program over a total sky area of 14,000 deg$^2$. 

Our forecasts here are based on the specifications of these spectroscopic samples as predicted through the Survey Validation campaign of DESI \cite{DESI_SV}. (1) The lowest-redshift sample that we consider (such that it has enough survey volume) consists of Luminous Red Galaxies (LRG) spread over the redshift range $0.4 < z < 1.1$. (2) We also consider Emission Line Galaxies (Low Priority targets) from the `ELG (LOP)' sample, that were selected as optimized targets over the intermediate redshift range: $1.1 < z < 1.6$, and (3) a high-redshift sample of discrete tracer quasars (QSO) spanning $1.6 < z < 2.1$. Thus, similar to the cosmological forecasts in \cite{DESI_SV}, we conservatively consider only the densest tracer within each redshift bin, so that each of the three samples are non-overlapping and can be used to test gravity at different cosmic times. 

\begin{table*}[t]
    \begin{tabular}{lcccc}
        \toprule
        & ~Redshift range~ & ~Effective redshift~ & ~Total number density~ & ~Survey Volume~ \\
        & $z$~ & ~$\zeff$~ &~in 2D (n deg$^{-2}$)~ & ~$V$ [Gpc]$^3$~ \\
        \midrule
        DESI LRG &0.4 - 1.1& 0.731 & 478 & 62.80\\
        DESI ELG (LOP)  &1.1 - 1.6& 1.325 & 452 & 75.41\\
        DESI QSO &1.6 - 2.1& 1.842 & 55 & 87.89\\
        \bottomrule
    \end{tabular}
    \caption{Relevant survey specifications for the three spectroscopic galaxy samples from the five-year main survey of the Dark Energy Spectroscopic Instrument (DESI) considered in this work.}
    \label{tab:DESI}
\end{table*}

Table \ref{tab:DESI} summarizes overall specifications for each of the DESI galaxy samples, including their associated $\zeff \equiv \zeff^{\kappa g}$, as defined in Section \ref{zeff}. Their redshift distributions ($\dndz$) assuming a bin-width of $\Delta z=0.1$, and their equivalent survey volume $V$, are calculated based on \cite{DESI_SV} (see Table 7 and Fig. 15 therein). The overlapping sky fraction $f_{\mathrm{sky}}$ of DESI galaxies with ACT and SO is taken to be around 0.19 \cite{JKim_ACTdr6_DESI_photo_LRGs_2024} and 0.23 \cite{SO}, respectively; the corresponding variance of $\kgCl$ and $\vgCl$ (through the survey volume $V$) gets scaled accordingly by a factor of (1/$f_{\mathrm{sky}}$). 

Since the $\hatVG$ estimator is constructed to measure the $E_G$ statistic only at sub-horizon linear scales, most of the involved quantities are straightforwardly computed in the linear regime as described above. However, the kSZ effect induces a squeezed $\langle Tgg\rangle$-type bispectrum (see Section \ref{ksz}), which leads to a quadratic estimator $\vrad(\mathbf{k})$ of the \textit{large-scale} radial velocity field, constructed from weighted $\langle Tg\rangle$ pairs at \textit{small} scales ($k_S\sim1 - 3$ Mpc$^{-1}$). Thus, to calculate the (scale-independent) noise power spectrum of $\vrad$ (Eq.~\eqref{Nvr}), we compute the required $P_{gg}(k_S)$ and $P_{ge}(k_S)$ power spectra at small scales within the halo model framework \cite{halomodel} using the code \texttt{hmvec}\footnote{https://github.com/simonsobs/hmvec} \cite{Smith2018}. We follow the same prescription as detailed in Appendix B of \cite{Smith2018}, where the HOD for the galaxy sample considered is estimated in \texttt{hmvec} based on its number density. For the purpose of our forecasts, we assume the simulations-based Battaglia electron density profile with `AGN' sub-grid feedback \cite{Battaglia2016} in our fiducial model (see Section \ref{ksz} and \ref{snr} for discussions of the potential impact of this choice on our results).

Now, given the wide redshift ranges spanned by the three DESI samples considered, we forecast $E_G$ constraints assuming a redshift-dependent linear galaxy bias model in each $z$-bin (of $\Delta z=0.1$) for them, as described in \cite{DESI_SV}: (1) $b^{\mathrm{LRG}}_g(z) = 1.7/D(z)$, (2) $b^{\mathrm{ELG}}_g(z) = 0.84/D(z)$, and (1) $b^{\mathrm{QSO}}_g(z) = 1.2/D(z)$. Here, $D(z)$ denotes the linear growth factor, which can be computed by integrating the corresponding linear growth rate $f(z)$ obtained from CAMB, and normalized such that $D(z=0) = 1$. While the expectation value of the $\hatVG$ estimator is expected to be insensitive to the linear galaxy bias $b_g$ up to first order, the redshift-evolution of $b_g$ within a galaxy sample slightly modifies the relative contributions from different $z$-bins while projecting along the LOS (Eqs.~\eqref{Kg},~\eqref{eqn:Clvg_def}). 

\subsection{Detection Significance}\label{snr}
\begin{figure*} 
  \centering
  \subfloat[\centering ACT$\times$DESI]{\includegraphics[width=0.48\textwidth]{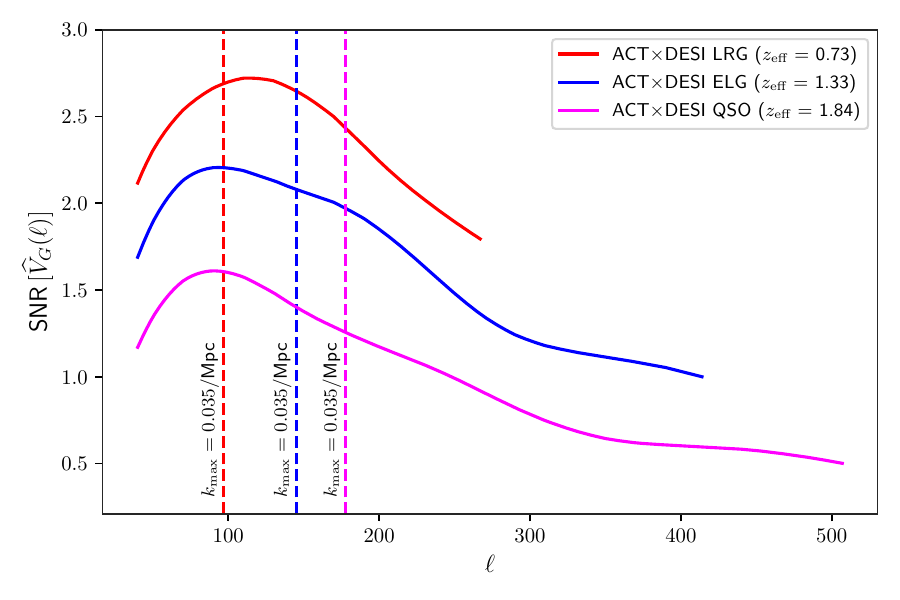}\label{fig:ACT_DESI_1}}
  \subfloat[\centering SO$\times$DESI]{\includegraphics[width=0.48\textwidth]{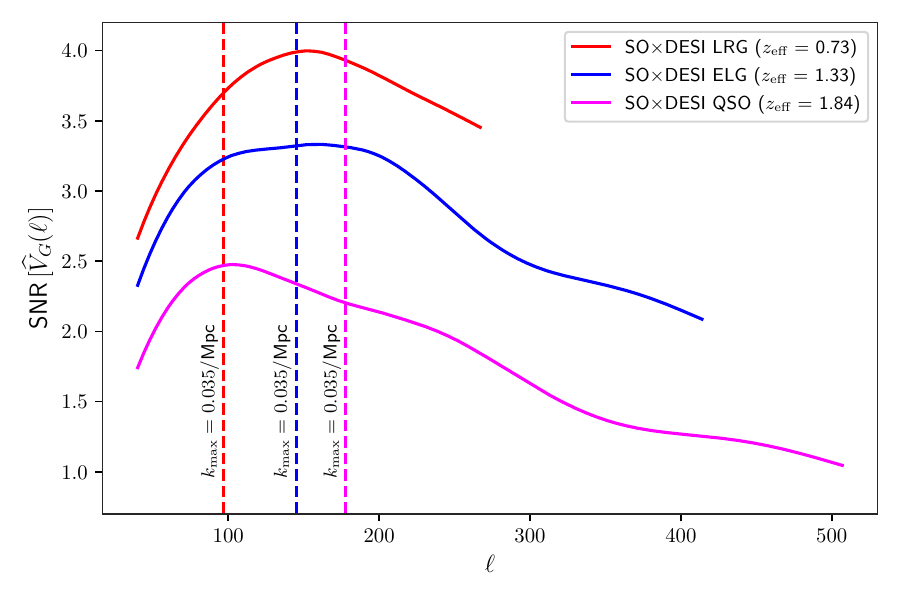}\label{fig:SO_DESI_1}}
  \caption{Forecasted signal-to-noise ratios as a function of multipole $(\mathrm{SNR}(\ell) \equiv E_G(\ell, \zeff)/\sigma\,[\hatVG(\ell, \zeff)])$ for measuring the $E_G$ statistic at $\zeff$  
 using the proposed $\hatVG$ estimator. Results for the three DESI galaxy samples considered (Section \ref{survey}): LRG, ELG, and QSO, are shown as red, blue, and pink curves, respectively, by combining them with CMB data (a) from ACT DR6, and (b) from SO. For each galaxy sample, the highest multipole we consider is given by $= (k_{\max}\chieff)$, with a default assumed value of $k_{\max}=0.1 \, \text{Mpc}^{-1}$ (the `squeezed' limit \cite{Smith2018}); a more stringent possible scale-cut \cite{giri2020} of $k_{\max}=0.035 \, \text{Mpc}^{-1}$ is also depicted here for reference by dotted lines.}
  \label{fig:SNR_ell}
\end{figure*}
We now present forecasts for measuring the $E_G$ statistic using the proposed $\hatVG$ estimator. Throughout, unless stated, we assume a $k_{\max}=0.1 \, \text{Mpc}^{-1}$ for the $\hatVG$ estimator, since this corresponds to the squeezed-limit regime in which the kSZ velocity-reconstruction method is established (Section \ref{ksz}), and its associated velocity bias (if any) is expected to be scale-independent. The maximum multipole considered for each survey combination (with an associated $\zeff$) is set to be $\ell_{\max}\approx (k_{\max}\chieff)$. Thus, assuming this scale-cut at the effective redshifts given in Table \ref{tab:DESI}, we are estimating the $E_G$ statistic at linear scales only (unlike the forecasted results from \cite{Pullen+2015_forecasts}, which extend to include quasi-linear scales too). 

For each combination, we consider one of the three DESI galaxy samples, along with CMB data from either ACT DR6 or SO, for obtaining both the CMB lensing ($\kgCl$) and the kSZ-reconstructed observables ($\vgCl$), which together constitute the $\hatVG$ estimator (Section \ref{estimator_deriv}). Following the method and specifications detailed in the previous subsection, we forecast signal-to-noise ratios (SNR) for measuring the $E_G$ statistic at $\zeff$ using $\hatVG$ as: $\mathrm{SNR}(\ell) \equiv E_G(\ell, \zeff)/\sigma\,[\hatVG(\ell, \zeff)]$, which are estimated assuming the fiducial $\Lambda$CDM+GR model. Figure \ref{fig:SNR_ell} shows the SNRs as a function of scale; for each data combination, the SNR peaks for multipoles in the sub-range $70 \lesssim \ell \lesssim 200$. 

We find that the overall noise covariance of $\hatVG$ is dominated by the error on $\kgCl$ (Eq.~\eqref{var-Kg}), which includes the galaxy shot noise, and which receives a large contribution from the lensing reconstruction noise. On the other hand, the noise of kSZ-reconstructed velocities ($N_{v_{r}}$; Eq.~\eqref{Nvr}) is expected to improve rapidly with such sensitive, high-resolution CMB data. 
As a cross-check of our numerical implementation, we also estimated the kSZ velocity-reconstruction noise ($N_{v_{r}}$) for SO and a preliminary DESI galaxy sample \cite{DESI_expt} that includes BGS galaxies along with LRG, ELG, and QSO galaxies, as considered in \cite{Smith2018}. Upon comparison, our estimated noise, $N_{v_{r}}$, reproduces their forecasted results (e.g. Fig. 5 of \cite{Smith2018}).   

For reference, in Fig.\,\ref{fig:SNR_ell}, we also show a possible scale-cut that conservatively limits the analysis to scales $k \lesssim 0.035 \,\text{Mpc}^{-1}$, as suggested in \cite{giri2020} to avoid any scale-dependence in the velocity bias ($b_v$), based on N-body simulations. While we discuss further forecasts with this stringent scale-cut in the Appendix, its exact value depends on particular survey specifications, and is likely too conservative/stringent for the high number density DESI galaxy samples that we have considered. Hence, for the rest of this work, we continue to assume $k \leq 0.1 \, \text{Mpc}^{-1}$, the squeezed-limit regime of kSZ tomography \cite{Smith2018, Munch}. 
\begin{table}[!h]
    \centering
    \begin{tabular}{lccc}
        \toprule
        & ~DESI LRG~ & ~DESI ELG LOP~ & ~DESI QSO~ \\
        \midrule
        ACT & 36 & 32 & 22 \\
        SO  & 56 & 55 & 39 \\
        \bottomrule
    \end{tabular}
    \caption{Cumulative SNRs of $\widehat{V}_G(\ell, \zeff)$ combined across all scales upto $k_{\max} = 0.1 \ \text{Mpc}^{-1}$ for different survey combinations of DESI galaxy samples and high-resolution CMB experiments.}
    \label{tab:SNR_1}
    \vspace{-20pt}
\end{table}
Now, since the fiducial GR prediction of the $E_G$ statistic is scale-independent, it can be estimated by combining measurements with the $\hatVG$ estimator across the entire scale range considered. Thus, we also compute cumulative SNRs for measuring the $E_G$ statistic at the $\zeff$ of each survey combination, by combining the corresponding $\mathrm{SNR}(\ell)$ in quadrature. Table \ref{tab:SNR_1} shows the forecasted cumulative SNRs. Given the much larger number densities of the DESI LRG and ELG samples, the cumulative SNRs of $\hatVG$ using them are a factor of $\sim 1.5$ times higher than those with the DESI QSO sample. 

Moreover, the cumulative SNR improves by a factor of $\sim 1.5-2$ with upcoming CMB data of higher sensitivity from SO as compared to ACT DR6, across all galaxy samples. Also, as discussed in Sections \ref{ksz} and \ref{estimator_deriv}, if the assumed fiducial model for the electron density profile \cite{Battaglia2016} differs from the true one (e.g. possibly due to higher levels of galactic feedback \cite{Hadzhiyska2024, BRGuechella2025, Hotinli2025}), the resulting kSZ velocity-reconstruction noise could be higher by a factor of up to $\sim 2$. However, since the lensing reconstruction noise dominates the overall error budget of $\hatVG$, this choice has a small ($<10\%$) impact on the overall cumulative SNRs quoted here (estimated here assuming a $b_v \approx 0.45$ as found in the latest kSZ-velocity reconstruction measurement \cite{Hotinli2025}). Importantly, even if a constant velocity bias $b_v\neq1$ is present in the reconstructed-velocities (after assuming a realistic baryon density profile), it would not alter the scale-dependence of the measured $E_G$ statistic. Thus, the proposed $\hatVG$ estimator can be used to robustly test the scale-independent prediction of GR at linear scales. 

\subsection{Distinguishing GR and Modified Gravity} 
\label{sub:GR_vs_MG}
\begin{figure*} 
  \centering
  \subfloat[\centering ACT$\times$DESI]{\includegraphics[width=0.48\textwidth]{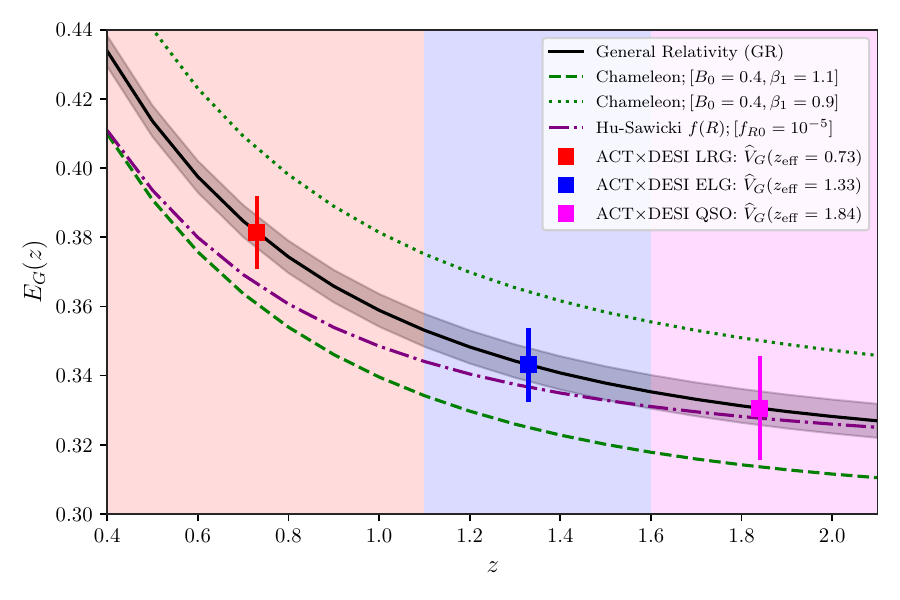}\label{fig:ACT_DESI_GR_MG}}
  \subfloat[\centering SO$\times$DESI]{\includegraphics[width=0.48\textwidth]{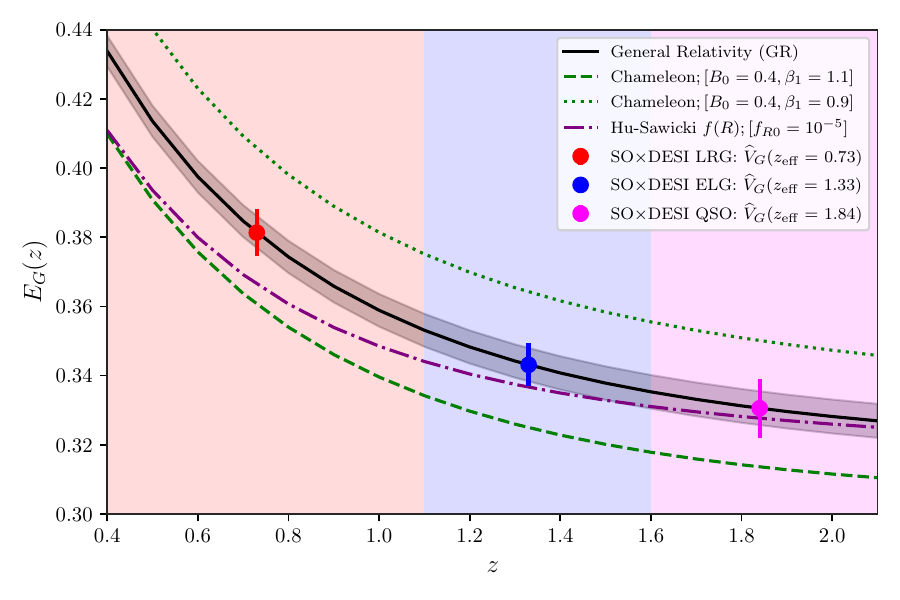}\label{fig:SO_DESI_GR_MG}}
  \caption{Predictions of the $E_G$ statistic as a function of redshift: from GR (black line), from the Hu-Sawicki $f(R)$ model ($f_{R0} = 10^{-5}$; violet curve), and from two representative Chameleon gravity models (green curves), obtained at each $z$ by averaging over the corresponding range of considered scales ($20 \leq \ell \lesssim k_{\max}\chi(z))$. The gray shaded region denotes the current associated uncertainty of the GR prediction. We also show error bars corresponding to the cumulative SNR of $\hatVG$ measurements using CMB data from (a) ACT DR6 and (b) SO, when combined with the DESI LRG (red), ELG (blue), and QSO (pink) galaxy samples, at their respective effective redshifts.}
  \label{fig:EG_z}
\end{figure*}

The results from the previous subsection show that the $\hatVG$ estimator can be used to robustly test the scale-independent prediction from GR of the $E_G$ statistic with high significance, where different galaxy samples probe this prediction at various cosmic times corresponding to their effective redshift. We now compare the $E_G$ statistic as predicted by GR versus a few specific modified gravity (MG) models, which were introduced in Section \ref{predic}. In this work, since we are probing gravity at linear, cosmological scales (where it is typically less tightly constrained, e.g. \cite{JLSchmidt_2016}), we consider the Hu-Sawicki $f(R)$ model \cite{HuSawfR2007} with a parameter value of $f_{R0} = 10^{-5}$ (and $n = 1$). Following \cite{Pullen+2015_forecasts}, we also consider a few representative models from the broader class of Chameleon gravity \cite{Khoury_prd_cham}, using the parameterization given by Eq.~\eqref{cham} with $s=4$, and values $(B_0, \beta_1)$ = (0.4, 1.1), (0.4, 0.9), or (3.2$\times 10^{-4}$, 1.2), as described in Section \ref{predic}. 

Figure \ref{fig:EG_z} shows the $E_G$ statistic's prediction from GR and the above MG models as a function of redshift (only), where the predictions at each redshift have been obtained by averaging over the entire scale range (with $k \leq 0.1$ Mpc$^{-1}$). The shaded gray region shows the estimated uncertainty of the GR prediction (Eq.~\eqref{EG_gr}), which is obtained by propagating current uncertainty on the $\Omega_{m, 0}$ parameter (assumed to be $\approx 0.005$ from current \textit{Planck}+ACT-DR6+DESI-DR1 BAO constraints). We compute the MG models' predictions (Eq.~\eqref{EG_mg}) under the $\mu-\gamma$ parametrization, by solving the differential equation (Eq.~\eqref{f_ODE}) for $f(k,z)$, which depends on $\mu(k, z)$. For comparison, we also show the 1$\sigma$ forecasted error bars corresponding to the cumulative SNR previously computed for each survey combination at its $\zeff$.  

While Figure \ref{fig:EG_z} averages over differences in the predicted scale-dependence of the $E_G$ statistic, it shows that a measurement with $\hatVG$ using the (five-year) DESI LRG galaxy sample and current (upcoming) CMB data from ACT DR6 (SO) would separate the overall $E_G(\zeff)$ value predicted by GR and by the considered MG models at an approximately $\gtrsim 1\sigma$ ($\gtrsim 2\sigma$) level. As expected, differences in growth of structure under GR and MG models are most pronounced at late times; thus, among the three galaxy samples we have considered, the lowest-redshift sample of DESI LRGs is the most crucial one for distinguishing between them. Although sensitive measurements with $\hatVG$ using the DESI ELG or QSO galaxy samples would serve as a robust test of the fiducial $\Lambda$CDM+GR model at earlier times, in the rest of this subsection, we only focus on the DESI LRG sample for statistically \emph{distinguishing} between gravity scenarios\footnote{For completeness, we note that there is another DESI sample \cite{DESI_SV} consisting of Bright Galaxies (BGS) that spans even lower redshifts: $0 \leq z \leq 0.4$. Following our numerical implementation, we forecast a cumulative SNR of around 8 for DESI BGS$\times$SO (as compared to $\sim 56$ for DESI LRG$\times$SO). However, we do not consider the BGS galaxy sample further in this work, since it has a much smaller survey volume, and has much lesser number of linear modes with $k\leq 0.1/$Mpc.}. 

\begin{table*}[!htb]
    \centering
    $\chi_{\mathrm{rms}} \equiv \sqrt{\chi^2_{\mg}}$ for Modified Gravity models \\[0.5em]
    \begin{tabular}{lccc}
        \toprule
        & ~~~~ Hu-Sawicki~~~~  &  ~~~~Chameleon ~~~~ & ~~~~Chameleon~~~~ \\
             & $f_{R0}\sim10^{-5}$ & $(B_0, \beta_1) = (0.4, 1.1)$~~~~ &  $(B_0, \beta_1) = (0.4, 0.9)$ \\
        \midrule
        ACT$\times$~DESI LRG~ & 1.19 & 1.96 & 2.28   \\
        SO$\times$~DESI LRG~ & 2.03 & 3.01 & 3.52   \\
        \bottomrule
    \end{tabular}
    \caption{Forecasted $\chi_{\mathrm{rms}} \equiv \sqrt{\chi^2_{\mg}}$ values (Eq.~\eqref{chi-mg}) quantifying the ability of future measurements with the $\hatVG$ estimator using DESI LRG galaxies and CMB data from ACT DR6 (top row) or SO (bottom row), to distinguish between GR and certain modified gravity models (Section \ref{predic}) at linear cosmological scales.}
    \label{tab:chisq_rms}
\end{table*}

\begin{figure}
  \centering
  {\includegraphics[width=0.49\textwidth]{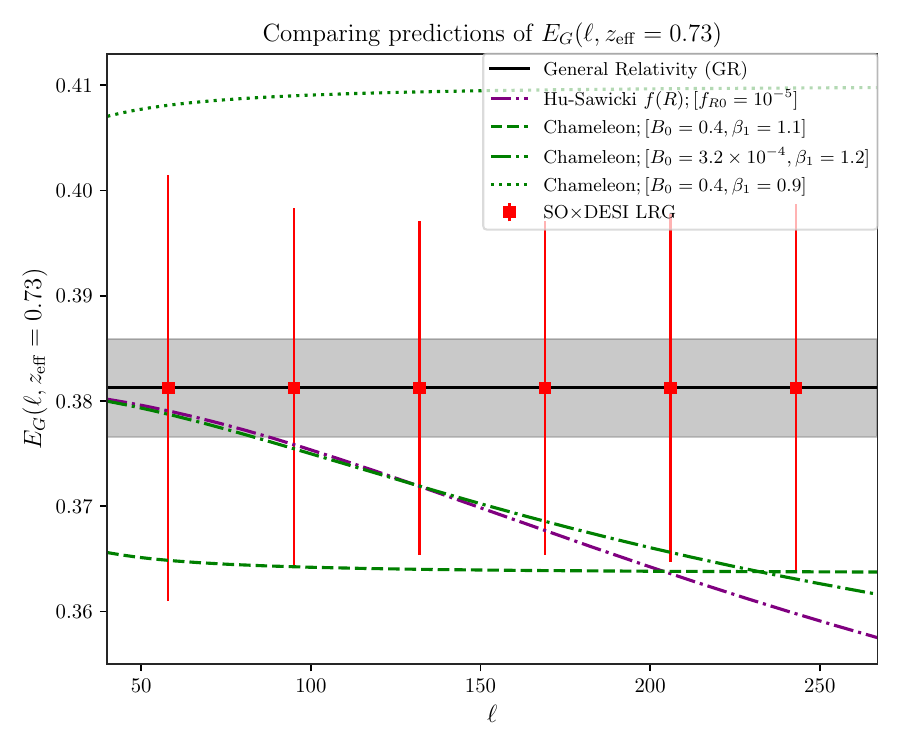}\label{fig:f1}}
  \caption{Comparing predictions of the $E_G$ statistic as a function of scale at $\zeff = 0.73$ (corresponding to the DESI LRG sample) from GR (black line), the Hu-Sawicki $f(R)$ model ($f_{R0} = 10^{-5}$; violet curve), and a few representative Chameleon gravity models (green curves). The gray shaded region denotes the current associated uncertainty of the GR prediction. 
  For reference, we also show the forecasted error bars of a $\hatVG$ measurement with SO$\times$DESI LRG (red) centered on the GR prediction, where the multipole range is split into six linearly spaced $\ell-$bins.}
  \label{fig:EG_vs_l_lrg}
  \vspace{-15pt}
\end{figure}
Figure \ref{fig:EG_vs_l_lrg} shows the GR prediction of the $E_G(\ell)$ statistic (at $\zeff= 0.73$ for DESI LRGs), along with the MG models' predictions, which are scale-dependent due to the corresponding $f(k, \zeff)$. For comparison, we also show the forecasted error bars for SO$\times$DESI LRG using the $\hatVG$ estimator (in red), by splitting the multipole range into six linearly spaced $\ell-$bins. We find that while predictions from the Chameleon MG models with $(B_0, \beta_1)$ = (0.4, 1.1) and (0.4, 0.9) are separated by $\gtrsim 1\sigma$ with respect to the GR prediction across $\ell-$bins, those from the Chameleon model with $(B_0, \beta_1)$ =(3.2$\times 10^{-4}$, 1.2) and the Hu-Sawicki $f(R)$ model are close to each other, and are distinguishable from GR only at the higher-$\ell$ end. However, we keep our analysis restricted to linear scales only ($k\leq0.1/$Mpc), where the overall measurement remains mostly insensitive to the (linear) galaxy bias.

We now quantify the ability of our proposed $\hatVG$ estimator to serve as a diagnostic test to distinguish between GR and MG models. We formalize this idea by setting it up as a binary hypothesis testing problem, where the fiducial $\Lambda$CDM+GR model is taken to be the null hypothesis ($H_0$) and a particular MG model is taken to be the alternative hypothesis ($H_1$). This problem can be solved using the Log Likelihood Ratio test, in which for a given set of observations at $\zeff$, $\widehat{V}_G^{\mathrm{obs}}(\ell, \zeff)$, we compute the test statistic:
\begin{align}
    \mathcal{L} = \sum_{\ell} \log \left( \frac{h_0(\widehat{V}_G^{\mathrm{obs}}(\ell, \zeff))}{h_1(\widehat{V}_G^{\mathrm{obs}}(\ell, \zeff))}\right).
\end{align}
In the above equation, $h_0$ and $h_1$ denote the probability density functions of the $E_G$ statistic under $H_0$ and $H_1$, respectively. The Log Likelihood Ratio, $\mathcal{L}$, is then compared to a carefully chosen threshold $\tau$; if $\mathcal{L} \geq \tau$, $H_0$ is declared to be true, otherwise $H_1$ is accepted.

In the absence of measured data, for the purpose of our forecasts, we can analytically quantify this approach by making a simplifying assumption that $\hatVG$ is a Gaussian. In particular, under $H_0$ (GR), we assume $\{V_G(\ell, \zeff)\}_{\ell}$ (i.e. as a function of $\ell$) are independent Gaussian random variables with an expected value and variance of $E_G^{\mathrm{GR}}(\ell, \zeff)$ (which is $\ell$-independent) and $\sigma^2[V_G(\ell, \zeff)]$, respectively. Similarly, under $H_1$ (MG), $\{V_G(\ell, \zeff)\}_{\ell}$ are taken to be independent Gaussian random variables with the same variance as under the fiducial $H_0$ (GR), but with mean equal to $E_G^{\mathrm{MG}}(\ell, \zeff)$. Under this setup, we define a key quantity that determines the efficacy of the $\hatVG$ method in distinguishing between an MG model and GR:
\begin{align}\label{chi-mg}
    \chi^2_{\mathrm{MG}} \equiv \sum_{\ell} \frac{(E_G^{\mathrm{MG}}(\ell, \zeff) - E_G^{\mathrm{GR}}(\zeff))^2}{\sigma^2[V_G(\ell, \zeff)]},
\end{align}
which is twice the expected value of the log-likelihood ratio ($\mathcal{L}$) under the null hypothesis (GR). In the above definition, $\zeff$ and the range of $\ell$ for the summation are determined by the specific survey combination. $\chi^2_{\mathrm{MG}}$ closely resembles the well-known $\chi^2$ quantity used in cosmological analyses (e.g. \cite{ModCosmo_Dodelson_2003, Pullen+2015_forecasts}).
The value of $\chi^2_{\mathrm{MG}}$ quantifies the efficacy of the $\hatVG$ estimator as a distinguishing test of gravity (GR or MG), by determining the range of possible values of type I and type II errors, i.e., $(\alpha, \beta)$, through the following relation:
\begin{align}
   \chi_{\mathrm{rms}} = \sqrt{\chi^2_{\mathrm{MG}}} \geq \mathcal{F}^{-1}(1- \alpha) + \mathcal{F}^{-1}(1-\beta). 
\end{align}
Here, $\mathcal{F}(\cdot)$ denotes the cumulative distribution function of the standard normal and $\mathcal{F}^{-1}$ denotes its inverse. $\alpha$ and $\beta$ denote the probabilities of false alarm (i.e., mistaken rejection of $H_0$ (GR), or the significance level) and of misdetection (i.e., mistaken failure to reject  $H_0$ (GR)), respectively. Thus, larger values of $\chi_{\mathrm{rms}}$ allow the errors $\alpha$ and $\beta$ to be simultaneously small, thereby enabling us to confidently choose one gravity model (or hypothesis) over the other based on the measured values of $\hatVG(\ell, \zeff)$. 

Table \ref{tab:chisq_rms} shows the estimated $\chi_{\mathrm{rms}}$ values (Eq.~\eqref{chi-mg}) for certain MG models considered in this work (under the $\mu-\gamma$ parametrization; see Section \ref{predic}). These values forecast the ability of measurements with the proposed $\hatVG$ estimator to distinguish between GR and MG models at the largest possible cosmological (linear) scales. For example, a $\chi_{\mathrm{rms}}$ value of $\geq 2$ allows for both type I ($\alpha$) and II ($\beta$) errors to be simultaneously less than 16$\%$. Similarly, a $\chi_{\mathrm{rms}} \geq 3$ allows for errors ($\alpha, \beta$) = (0.05, 0.1). Hence, a $\hatVG$ measurement with ACT DR6$\times$DESI LRG (SO$\times$DESI LRG) can distinguish between GR and the two Chameleon gravity models (the Hu-Sawicki $f(R)$ model) with 84$\%$ confidence. Moreover, an SO$\times$DESI LRG measurement will correctly declare GR over the two Chameleon gravity models with 95$\%$ confidence ($\alpha = 0.05$). At the same time, it can correctly detect the Chameleon ($\beta_1 = 1.1$) model and the Chameleon ($\beta_1 = 0.9$) model with 90$\%$ and 95$\%$ confidence levels, respectively.

\section{Future Outlook and Conclusions}
\label{sec:conclusion}
The $E_G$ statistic is a powerful tool for distinguishing between dark energy and modified gravity theories as two explanations for the accelerated expansion of the Universe. In this work, we have constructed a novel $\hatVG$ estimator to measure the $E_G$ statistic, by combining cross-correlations of spectroscopic galaxy samples with CMB lensing convergence and kSZ-reconstructed velocities. While all previous measurements and estimators of the $E_G$ statistic have relied on Redshift Space Distortions (RSD), the $\hatVG$ estimator instead uses kSZ tomography to access the velocity information. Aside from being a complementary approach to RSD for extracting velocities, kSZ velocity-reconstruction can constrain the largest scale modes with lower noise than RSD and the galaxy survey it uses \cite{Smith2018, Munch}. 

Equipped with this new estimator, we can perform a first measurement with publicly available ACT-DR6 \cite{ACTDR6_Qu2024} and DESI-DR1 spectroscopic LRG \cite{desi_dr1} data already at hand. One component of the $\hatVG$ estimator, the angular galaxy-CMB lensing cross-power spectrum, has recently been measured with these datasets \cite{MausWhite_2025_ACT6_specDESI_LRG}. To incorporate and interpret such a measurement accurately, we would need to account for possible additional systematics, such as errors at linear scales induced due to uncertainties in the halo occupation distribution of the considered galaxies. 
Also, while we assumed a simplified snapshot geometry in the kSZ velocity-reconstruction framework \cite{Smith2018, Lague2024} for our forecasts, the \texttt{kszx} package\footnote{https://github.com/kmsmith137/kszx} for this method additionally accounts for sky-cuts, masking, and evolution along the light cone, as recently demonstrated in \cite{Hotinli2025}. We plan to build a pipeline based on this code to perform kSZ velocity reconstruction using the same datasets, for the other component of our proposed estimator. 

While obtaining the quadratic estimator of kSZ-reconstructed velocities (Section \ref{ksz}), a more realistic baryon density profile (informed by recent kSZ-stacking measurements using DESI data \cite{Hadzhiyska2024, BRGuechella2025, Hotinli2025} or by appropriately-calibrated hydrodynamical simulations \cite{Bigwood_2025}) could be assumed; this would limit the associated optical depth degeneracy. Importantly, even if a constant velocity bias $\neq 1$ is present in the kSZ measurement, this would have a $<10\%$ impact on the cumulative detection significances, and it would not alter the \emph{scale-dependence} of the measured $E_G$ statistic. Thus, measurements with the proposed $\hatVG$ estimator here can be used to robustly test the \emph{scale-independent prediction of GR} at linear scales.

As detailed in Section \ref{est}, we defined the precision $\hatVG$ estimator by combining $\kgCl$ with $\vgCl$, a kSZ-derived quantity. $\vgCl$ is obtained by appropriately projecting the 3D galaxy-velocity cross-power spectrum from kSZ velocity-reconstruction, and includes a suitable reweighing of the galaxy sample to match the effective redshifts of the observables. Based on this estimator, we present forecasts for measuring the $E_G$ statistic using ACT DR6 (current) and SO (upcoming) CMB maps, in combination with the LRG, ELG, and QSO spectroscopic samples from the main survey of DESI (spanning 5 years). The forecasted cumulative detection significances (combined across scales) are in the range $S/N \sim 20-55$, as shown in Section \ref{fore}. To accurately interpret such measurements with the $\hatVG$ estimator, it would be valuable for future works to estimate its covariance matrix using simulated maps \cite{LWenzl+2024_PlanckBOSS, LWenzl+2025_ACTBOSS}, to validate the analytical approximation adopted here, and quantify the off-diagonal terms. 

As further data of DESI galaxies becomes available, the detection significance of the corresponding $\hatVG$ measurements are expected to increase and approach the levels forecasted in this work. Moreover, with the SO LAT now online, and enhanced SO on the horizon \cite{ASO_2025}, its CMB convergence maps with improved lensing reconstruction noise would allow increasingly sensitive measurements of the $E_G$ statistic with this estimator. In this high-SNR regime, each of the spectroscopic DESI galaxy samples could be split further into smaller tomographic bins (e.g., as done in \cite{MausWhite_2025_ACT6_specDESI_LRG}). This would lead to stringent consistency tests of the concordant $\Lambda$CDM+GR model across various effective redshifts. Moreover, 
we forecast that upcoming SO$\times$LRG measurements with the proposed estimator will be able to distinguish between GR and certain MG models with high confidence. Thus, the novel approach presented in this work would allow us to leverage increasingly sensitive measurements of CMB secondaries in the near future to probe the nature of gravity at the largest observable scales.

\begin{acknowledgments}
We are grateful to Mathew Madhavacheril, Alex Lague, and Selim Hotinli for useful discussions regarding the kSZ velocity-reconstruction method. We thank Emmanuel Schaan, Bernardita Ried Guachalla, J. Colin Hill, and Gerrit Farren for their helpful comments. RP thanks Sudeep Salgia for useful discussions, particularly regarding the log-likelihood ratio test, which contributed to Section \ref{sub:GR_vs_MG}. NB acknowledges the additional support from NASA grants 80NSSC18K0695 and 80NSSC22K0410. The work of RB is supported by NSF grant AST-2206088, NASA grant 22-ROMAN11-0011, and NASA grant 12-EUCLID12-0004.
\end{acknowledgments}

\bibliography{library}

\appendix
\section{Results with a stringent scale-cut}
\begin{figure*}[!htb] 
  \centering
  \subfloat[\centering ACT$\times$DESI]{\includegraphics[width=0.48\textwidth]{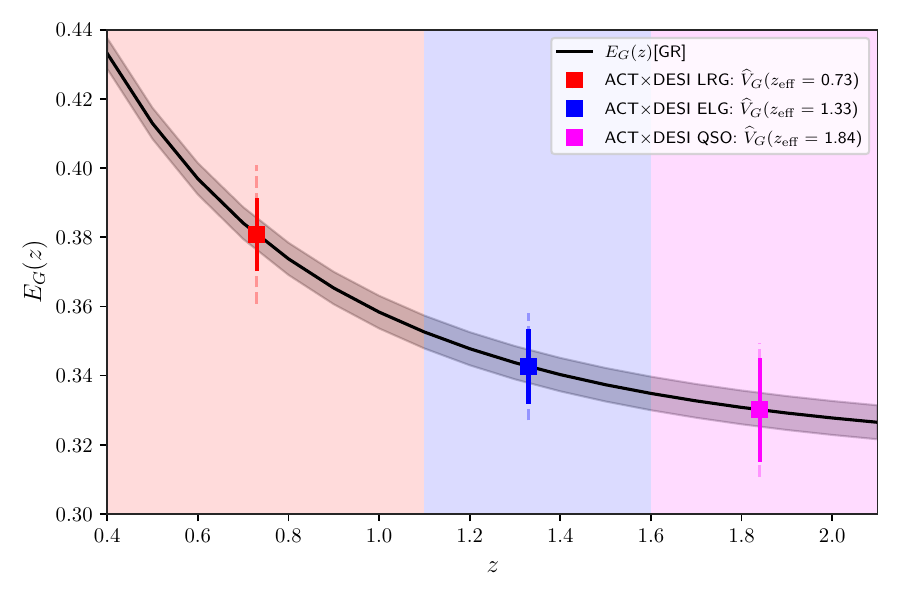}\label{fig:ACT_DEST_35}}
  \subfloat[\centering SO$\times$DESI]{\includegraphics[width=0.48\textwidth]{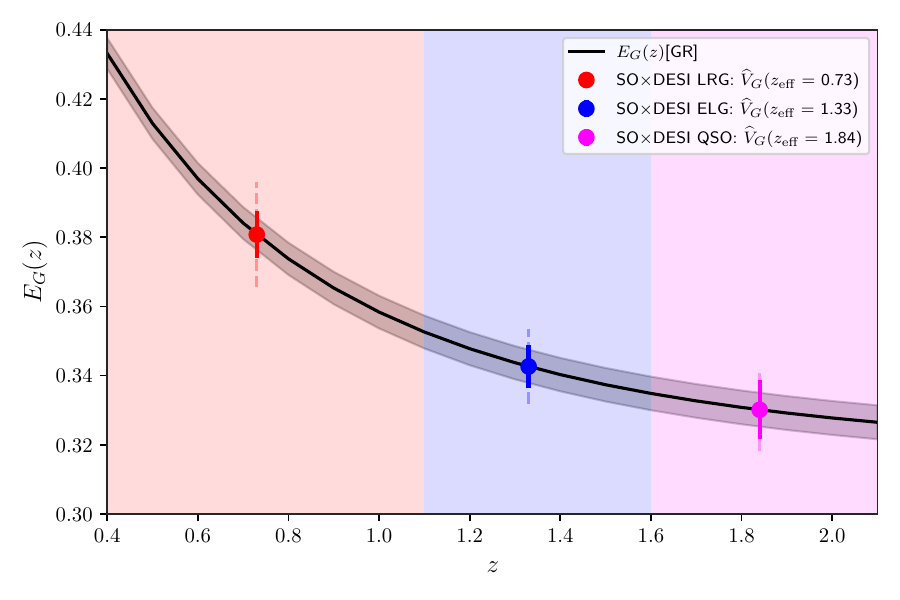}\label{fig:SO_DESI_35}}
  \caption{Prediction of the $E_G$ statistic as a function of redshift from GR (black line), obtained at each $z$ by averaging over the corresponding range of considered scales ($20 \leq \ell \lesssim k_{\max}\,\chi(z))$). The gray shaded region denotes the current associated uncertainty of the GR prediction. We show error bars corresponding to the cumulative SNR of $\hatVG$ measurements using CMB data from (a) ACT DR6 and (b) SO, when combined with the DESI LRG (red), ELG (blue), and QSO (pink) galaxy samples, at their respective effective redshifts. The solid error bars are computed with a scale-cut of $k_{\max}= 0.1 \ \text{Mpc}^{-1})$ (as shown in Fig. \ref{fig:EG_z} too). These are compared against the larger, dashed error bars (and lower cumulative SNR) that are obtained if a more stringent scale-cut of $k_{\max} = 0.035 \ \text{Mpc}^{-1}$ is used instead.} 
  \label{fig:EG_z_2}
\end{figure*}
The analysis and forecasts throughout this work assume a scale-cut of $k \leq 0.1 \, \text{Mpc}^{-1}$, which only considers linear modes that are within the squeezed-limit regime of kSZ velocity-reconstruction \cite{Smith2018, Munch}. A more stringent scale-cut of $k \lesssim 0.035\, \text{Mpc}^{-1}$ has also been suggested \cite{giri2020} based on N-body simulations, to avoid any possible scale-dependence induced in the velocity bias. While the exact value of this scale-cut depends on particular survey specifications, and is likely too conservative/stringent for the high number density DESI galaxy samples that we have considered, we discuss forecasts with a $k_{\max} = 0.035 \ \text{Mpc}^{-1}$ instead, for completeness.

Comparing Tables \ref{tab:SNR_1} and \ref{tab:SNR_2}, we find that using the much more conservative scale-cut of $k_{\max} = 0.035 \ \text{Mpc}^{-1}$ reduces the forecasted cumulative SNRs by a factor of $\sim 2$ for the DESI LRG$\times$ACT and DESI LRG$\times$SO measurements. The impact of such a stringent scale-cut is lesser for the ELG and QSO DESI samples (a factor of $\sim 1.7$ and $\sim 1.4$, respectively), since there is a relatively smaller loss of linear-scale $\ell$-modes at these deeper effective redshifts. Fig. \ref{fig:EG_z_2} shows the inflated error bars (dashed) and lower cumulative SNRs by using $k_{\max} = 0.035 \ \text{Mpc}^{-1}$. These can be compared against the associated uncertainty on the GR prediction (gray shaded region), which arises due to the propogated uncertainty on the $\Omega_{m,0}$ parameter (assumed to be $\approx 0.005$ based on current constraints \cite{DESI2024_cosmo}).

\begin{table}[b]
    \centering
    \begin{tabular}{|c|c|c|c|}
    \hline
         &  DESI LRG & DESI ELG LOP & DESI QSO\\
         \hline
         ACT & 19 & 22 & 17 \\
         \hline
         SO &  25 & 32 & 28 \\
         \hline
    \end{tabular}
    \caption{Cumulative SNRs of $\widehat{V}_G(\ell, \zeff)$ combined across all scales up to $k_{\max} = 0.035 \ \text{Mpc}^{-1}$ for different survey combinations of DESI galaxy samples and high-resolution CMB experiments.}
    \label{tab:SNR_2}
\end{table}

With the first kSZ-velocity reconstruction measurements performed recently using ACT data \cite{Lague2024, FMcCarthy2024, Hotinli2025, ALaiMunch2025}, it would be timely to conduct further tests of this method on simulations. These could be done with realistic mock galaxy catalogs representing the high number density of DESI spectroscopic galaxies, to determine a more suitable, validated scale-cut that would ensure a scale-independent velocity bias in such measurements. From the results above (Fig. \ref{fig:EG_z_2} and Table \ref{tab:SNR_2}), we conclude that measurements with the $\hatVG$ estimator would be a novel, robust test of the concordant $\Lambda$CDM+GR model at different effective redshifts, even with a stringent scale-cut of $k \leq 0.035\, \text{Mpc}^{-1}$. The forecasted cumulative SNR with SO in this scenario remains $>25$ for each of the three DESI galaxy samples considered in this work.  

\end{document}